\documentclass[prd,nofootinbib,notitlepage,10pt,aps]{revtex4-1}
\usepackage{epsfig,amsmath,amssymb,slashed}
\usepackage[utf8]{inputenc}
\usepackage{hyperref}
\usepackage{amsthm}
\usepackage{xcolor}

\theoremstyle{definition}
\newtheorem{theorem}{Theorem}
\newtheorem*{definition}{Definition}

\newcommand{\bra}[1]{\langle #1|}
\newcommand{\ket}[1]{|#1\rangle}

\newcommand{\media}[1]{\langle #1 \rangle}
\newcommand{\wad}[1]{\widehat a^\dagger(#1)}
\newcommand{\wa}[1]{\widehat a(#1)}
\newcommand{\wbd}[1]{\widehat b^\dagger(#1)}
\newcommand{\wb}[1]{\widehat b(#1)}
\newcommand{\di}{{\rm d}}
\newcommand{\Tr}{{\rm Tr}}
\newcommand{\ii}{i}
\newcommand{\dist}{{\rm dist}}

\newcommand{\h}[1]{\widehat{#1}}

\def\wT{{\widehat T}}

\def\wQ{{\widehat Q}}
\def\wP{{\widehat P}}
\def\wJ{{\widehat J}}

\def\wpsi{{\widehat{\psi}}}
\def\wrho{{\widehat{\rho}}}

\def\codevmu{{\stackrel{\leftrightarrow}{\partial^\mu}}}
\def\codevnu{{\stackrel{\leftrightarrow}{\partial^\nu}}}

\newcommand{\e}{{\rm e}}

\newcommand{\omegav}{\boldsymbol{\omega}}

\newcommand{\p}{{\rm p}}

\newcommand{\de}{\partial}
\newcommand{\be}{\begin{equation}}
\newcommand{\ee}{\end{equation}}                                                                               
\newcommand{\bea}{\begin{eqnarray}}
\newcommand{\eea}{\end{eqnarray}}

\begin{document}

\title{Exact equilibrium distributions in statistical quantum field theory 
with rotation and acceleration: scalar field}

\author{F. Becattini}\affiliation{Universit\`a di 
 Firenze and INFN Sezione di Firenze, Florence, Italy} 
\author{M. Buzzegoli}\affiliation{Universit\`a di 
 Firenze and INFN Sezione di Firenze, Florence, Italy}
 \author{A. Palermo}\affiliation{Universit\`a di 
 Firenze and INFN Sezione di Firenze, Florence, Italy}

\begin{abstract}
We derive a general exact form of the phase space distribution function and the thermal
expectation values of local operators for the free quantum scalar field at equilibrium 
with rotation and acceleration in flat space-time without solving field equations in 
curvilinear coordinates. After factorizing the density operator
with group theoretical methods, we obtain the exact form of the phase space distribution 
function as a formal series in thermal vorticity through an iterative method and we calculate 
thermal expectation values by means of analytic continuation techniques. We separately discuss the 
cases of pure rotation and pure acceleration and derive analytic results for the stress-energy 
tensor of the massless field. The expressions found agree with the exact analytic solutions 
obtained by solving the field equation in suitable curvilinear coordinates for the two cases 
at stake and already - or implicitly - known in literature. In order to extract finite values 
for the pure acceleration case we introduce the concept of analytic distillation of a 
complex function. For the massless field, the obtained expressions of the currents are 
polynomials in the acceleration/temperature ratios which vanish at $2\pi$, in full accordance 
with the Unruh effect. 
\end{abstract}

\maketitle

\section{Introduction}
\label{intro}

The goal of thermal quantum field theory is to calculate mean values of physical quantities 
at thermodynamic equilibrium, henceforth denoted as thermal expectation values. This is a well 
known subject for the familiar thermodynamic equilibrium, described by the grand-canonical 
ensemble density operator:
\be\label{homo1}
 \wrho = \frac{1}{Z} \exp \left[- \widehat H/T_0 + \mu_0 \wQ/T_0\right],
\ee
$T_0$ being the temperature and $\mu_0$ the chemical potential coupled to a conserved charge $\wQ$, 
and $Z$ the partition function. 

However, this is not the only possible form of global\footnote{We use global to distinguish it 
from local. The word global implies an actual equilibrium, with a density operator depending only 
on conserved quantities and with vanishing entropy production.} thermodynamic equilibrium in 
special relativity. It has become a topic of interest the calculation of thermal expectation values 
for the general form of global thermodynamic equilibrium in flat space-time, which is described by 
the density operator:
\be\label{general}
  \wrho = \frac{1}{Z} \exp \left[ - b_\mu {\wP}^\mu  
  + \frac{1}{2} \varpi_{\mu\nu} \wJ^{\mu\nu} + \zeta \wQ \right],
\ee
where the $\wP^\mu$ are the total four-momentum operators and $\wJ^{\mu\nu}$ the total angular-momentum 
boost operators; $b$ is a timelike four-vector and $\varpi$ an antisymmetric constant tensor which is 
known as {\em thermal vorticity}. Its form shows that the most general global equilibrium density operator 
in Minkowski spacetime involves all 10 generators of its maximal symmetry group, the proper ortochronous 
Poincar\'e group, with 10 constant coefficients. The operator \eqref{general} and its derivation from the 
generally covariant form as well as the relations between thermodynamic and hydrodynamic fields and the 
parameters $b,\,\varpi$ have been discussed in detail elsewhere~\cite{Becattini:2012tc,Becattini:2014yxa,Becattini:2015nva,Buzzegoli:2018wpy,Becattini:2019dxo}. 
Here we just recollect that \eqref{general} implies that the four-temperature field $\beta$ is given by:
\be\label{fourtemp}
 \beta_\mu = b_\mu + \varpi_{\mu\nu} x^\nu,
\ee
whose amplitude $\sqrt{\beta^2}$ is the inverse of the local comoving temperature $T$ and its
direction defines a four velocity $u$. The thermal vorticity can be decomposed into two space-like fields 
as:
\be\label{thvort}
  \varpi_{\mu\nu} = \epsilon_{\mu\nu\rho\sigma} w^\rho u^\sigma +
  	(\alpha_\mu u_\nu - \alpha_\nu u_\mu)
\ee
where, in the case of global equilibrium only:
\be\label{alphaw}
  \alpha^\mu = \frac{A^\mu}{T},  \qquad \qquad  w^\mu = \frac{\omega^\mu}{T},
\ee
$A^\mu$ being the acceleration field and $\omega^\mu$ the kinematic vorticity field.

The calculation of thermal expectation values of quantities such as the stress-energy tensor, conserved 
charge currents or spin density matrices with the operator \eqref{general} has a theoretical interest
in the framework of statistical quantum field theory in general spacetimes
~\cite{Gransee:2015aba,Panerai:2015xlr,Fredenhagen:2018obu} but it is also phenomenologically relevant 
for relativistic fluids undergoing very strong accelerations and rotations, such as the QCD plasma in 
high energy nuclear collisions \cite{STAR:2017ckg}. Indeed, with $b$ and $\varpi$ depending on the 
space-time point $x$, the operator \eqref{general} is the lowest order approximation of the local 
thermodynamic equilibrium density operator~\cite{zubarev,weert,Becattini:2014yxa,hongo,Harutyunyan:2018cmm,Becattini:2019dxo} which is 
needed to calculate those mean values in relativistic fluids. 

For free boson fields, it can be expected that any thermal expectation values arises from the Bose-Einstein 
distribution function in its covariant form (Juttner distribution):
\be\label{disfunc}
 f(x,p) = \frac{1}{\e^{\beta \cdot p - \zeta q}-1}
\ee
with $\beta$ in \eqref{fourtemp} including all the effects of acceleration and rotation. In fact,
it is known that a non-vanishing thermal vorticity implies the appearance of quantum corrections to the
quantities obtained by integrating the \eqref{disfunc}. They have been calculated perturbatively 
\cite{Becattini:2015nva,Buzzegoli:2018wpy,Prokhorov:2019cik,Prokhorov:2019hif,Prokhorov:2019yft} 
for small values of the adimensional thermal vorticity $\varpi$ (or, equivalently, small values of the 
ratios \eqref{alphaw}) with the operator expansion technique or the functional approach \cite{Kovtun:2016lfw,Kovtun:2018dvd}. 
However, not much is known about the full global equilibrium expression except for the two special 
cases of pure acceleration \cite{Becattini:2017ljh,Becattini:2019poj} and rotation \cite{AmbrusScalar} 
where the exact solution is obtained by solving field equations in suitable curvilinear coordinates. 
Besides the theoretical value, the knowledge of exact solutions can be of relevance for the physics 
of the QCD plasma in that the ratios \eqref{alphaw} can locally be ${\cal O}(1)$, according to the 
hydrodynamic simulations, and imply significant corrections to the leading order expression obtained 
in the perturbative expansion. The knowledge of the exact solution of the scalar field is also 
the first crucial step to derive an exact solution for the Dirac field. This can be of 
great importance to derive a complete expression of the fermion polarization in a relativistic 
fluid, a problem which has lately become phenomenologically very relevant.

In this work, we present a method to calculate the exact expression of the distribution function 
in statistical quantum field theory in global thermodynamic equilibrium with rotation and acceleration
without solving the field equations in curvilinear coordinates, 
that is the exact form of \eqref{disfunc} with all quantum corrections. The method is based on the
factorization of \eqref{general} and an iterative procedure to determine the thermal expectation values 
of creation and annihilation operators for imaginary thermal vorticity. The physical solutions are 
thereafter obtained extracting the analytic part of the formal series found, a mathematical procedure 
introduced in this work and named {\em analytic distillation}, followed by an analytic continuation to real 
thermal vorticity. We then compare the obtained results with those calculated by solving field 
equations in two special cases - pure acceleration and pure rotation - and find complete agreement. 
Finally, we demonstrate the power of this method by calculating the thermal expectation
value of the field squared in a newly explored global equilibrium case where both the acceleration 
and the vorticity vectors are non-vanishing.

The paper is organized as follows: in Section \ref{method} we introduce the factorization of the density 
operator and, after continuing to imaginary thermal vorticity, we iteratively compute the expectation 
value of creation and annihilation operators. In Sections \ref{phasespace} and \ref{properties} we 
determine the exact form of the distribution function and study its properties. In Section \ref{math},
we introduce the operation of \emph{analytic distillation} which is necessary to obtain the exact 
results for physical acceleration and rotation presented in Sections \ref{sec:accel} and \ref{rotation} 
respectively for the massless scalar free field. Finally, in Section \ref{newresult}, a new result is 
derived for the free massless scalar field at equilibrium with both acceleration and rotation.

\subsection*{Notations}
In this paper we adopt the natural units, with $\hbar=c=K=1$.  
The Minkowskian metric tensor $g$ is ${\rm diag}(1,-1,-1,-1)$; for the Levi-Civita
symbol we use the convention $\epsilon^{0123}=1$.  
We will use the relativistic notation with repeated indices assumed to 
be saturated. However, contractions of a single index, e.g. $\beta_\mu p^\mu$ will be
sometimes denoted with a dot, i.e. $\beta \cdot p$; similarly, the contraction of two
indices, e.g. $\varpi_{\mu\nu} J^{\mu\nu}$ will be sometimes denoted by a colon, i.e.
$\varpi : J$.   
Operators in Hilbert space will be denoted by an upper hat, e.g. $\widehat O$.

\section{Thermal expectation values of annihilation and creation operators}
\label{method}

For free fields, the building block to calculate any statistical quantity (mean values, correlations) 
is the mean value of the combination of one creation and one annihilation operator of four-momentum 
eigenstates:
\be\label{mainmean}
 \Tr \left( \wrho \; \wad{p} \, \wa{p^\prime} \right).
\ee
Its value depends, of course, on the density operator; if $\wrho = (1/Z) \exp[-\widehat{H}/T]$ 
the result is a well known one - the familiar Bose-Einstein and Fermi-Dirac distributions - and the 
derivation is based on the use of the known commutation relations between $\wrho$ and $\h{a},\,\h{a}^\dagger$. 
However, if the density operator is \eqref{general}, finding an exact expression is not trivial. 
Before setting out to do that, it should be first pointed out that any other combination of creation 
or annihilation operators, such as 
$\media{\wad{p}\,\wad{p^\prime}},\media{\wa{p}\,\wa{p^\prime}}$ or combination of creation/annihilation
operators of particles and antiparticles in case of a charged scalar field, will vanish. This
happens because the density operator \eqref{general}, involving just the generators of Lorentz
transformation, translations and charge, does not change the number of particles and antiparticles. 
Hence:
\be\label{nulli}
\media{\wad{p}\,\wad{p^\prime}}= \media{\wa{p}\,\wa{p^\prime}} = \media{\wbd{p}\,\wbd{p^\prime}}
   = \media{\wb{p}\,\wb{p^\prime}} = \media{\wa{p}\,\wb{p^\prime}} = \media{\wa{p}\,\wbd{p^\prime}} 
   = 0.
\ee

In this section we will present a general method to calculate \eqref{mainmean} with the 
density operator \eqref{general}. We will first set $\varpi$ imaginary and factorize the 
density operator \eqref{general} by using a group theoretical method; then we will calculate 
\eqref{mainmean} for imaginary $\varpi$ by means of an iterative method and give the general 
solution in terms of a uniformly convergent series. Finally, we will discuss the analytic 
continuation to real $\varpi$.

\subsection{Factorization of the density operator}

As has been mentioned, we take advantage of Poincar\'e group relations which make it possible to 
factorize the density operator \eqref{general} into the product of two independent operators.

Let us start with the following very simple observation concerning the composition of translations 
and Lorentz transformations in Minkowski space-time. Let $x$ be a four-vector and apply the 
combination 
$$
{\sf T}(a) \, {\sf \Lambda} \, {\sf T}(a)^{-1},
$$
${\sf T}(a)$ being a translation of some four-vector $a$ and ${\sf \Lambda}$ a 
Lorentz transformation. The effect of the above combination on $x$ reads:
$$
 x \mapsto x - a \mapsto {\sf \Lambda}(x-a) \mapsto {\sf \Lambda}(x-a)+a
 = {\sf \Lambda}(x) + ({\sf I-\Lambda})(a)= {\sf T}(({\sf I-\Lambda})(a))
  ({\sf \Lambda}(x)).
$$ 
Since $x$ was arbitrary, we have:
$$
 {\sf T}(a) \, {\sf \Lambda} \, {\sf T}(a)^{-1} = {\sf T}(({\sf I-\Lambda})(a))
 {\sf \Lambda}.
$$
This relation has a representation of unitary operators in Hilbert space, which can be 
written in terms of the generators of the Poincar\'e group:
\be\label{poinc1}
\exp[i a \cdot \widehat P] \exp[-\ii \phi : \wJ/2] \exp[-i a \cdot \widehat P]
= \exp[\ii(({\sf I-\Lambda})(a))\cdot \widehat{P}] \exp[-\ii \phi : \wJ/2] ,
\ee
where $\phi$ are the parameters of the Lorentz transformation. By taking $\phi$ infinitesimal, 
we can obtain a known relation about the effect of translations on angular momentum operators:
$$
\exp[i a \cdot \widehat P] \wJ_{\mu\nu} \exp[-i a \cdot \widehat P]
=  \widehat{\sf T}(a) \wJ_{\mu\nu} \widehat{\sf T}(a)^{-1} = \wJ_{\mu\nu}
 - a_\mu \widehat P_\nu + a_\nu \widehat P_\mu \, .
$$
The left hand side of \eqref{poinc1} can now be worked out by using the above
relation:
\begin{align}\label{poinc2}
 & \exp[i a \cdot \widehat P] \exp[-\ii \phi : \wJ/2] \exp[-i a \cdot \widehat P]
 = \widehat{\sf T}(a) \exp[-\ii \phi : \wJ/2] \widehat{\sf T}(a)^{-1} \nonumber \\
& = \exp[-\ii \phi : \widehat{\sf T}(a) \wJ \widehat{\sf T}(a)^{-1}/2]
 = \exp[-\ii \phi : (\wJ - a \wedge \widehat{P})/2] = \exp[\ii \phi_{\mu\nu} a^\mu 
 \widehat{P}^\nu -\ii \phi_{\mu\nu} \wJ^{\mu\nu}/2] .
\end{align}
Hence, combining \eqref{poinc2} with \eqref{poinc1}, we have obtained the factorization:
\be\label{intermed}
\exp[\ii \phi_{\mu\nu} a^\mu \widehat{P}^\nu -\ii \phi_{\mu\nu} \wJ^{\mu\nu}/2] 
= \exp[\ii(({\sf I-\Lambda})(a))\cdot \widehat{P}] \exp[-\ii \phi : \wJ/2].
\ee
Now, since it holds
\be\label{poinc3}
 \ii({\sf I-\Lambda})(a) = \ii a - \ii 
 \sum_{k=0}^\infty \frac{(-\ii)^k}{2^k k!}(\phi: {\sf J})^k (a)
  = - \ii \sum_{k=1}^\infty \frac{(-\ii)^k}{2^k k!}(\phi: {\sf J})^k (a),
\ee
by setting:
$$
    V_\mu \equiv \ii \phi_{\mu\nu} a^\nu
$$
and taking into account that:
$$
 ({\sf J}_{\mu\nu})^\alpha_{\, \beta} = \ii \left( \delta^\alpha_\mu g_{\nu\beta} -
 \delta^\alpha_\nu g_{\mu\beta} \right),
$$
we have:
$$
 (\phi : {\sf J})(a)_\alpha = 2 \ii \phi_{\alpha\beta}a^\beta = 2V_\alpha.
$$
Therefore, the right hand side of the eq.~\eqref{poinc3} becomes:
$$
 - \ii \sum_{k=1}^\infty \frac{(-\ii)^k}{2^k k!}(\phi: {\sf J})^k (a)
  = - \ii \sum_{k=1}^\infty \frac{(-\ii)^k}{2^{k-1} k!}(\phi: {\sf J})^{k-1}(V)
  = - \sum_{k=0}^\infty \frac{(-\ii)^{k}}{2^{k} (k+1)!}(\phi: {\sf J})^{k}(V).
$$
Finally, the eq.~\eqref{intermed} becomes:
\be\label{intermed2}
\exp[- V \cdot \widehat{P} -\ii \phi : \wJ/2] 
= \exp[ - \tilde V (\phi) \cdot \widehat{P}] \exp[-\ii \phi : \wJ/2],
\ee
where  we denoted
$$
\tilde V (\phi)_\mu \equiv \sum_{k=0}^\infty \frac{(-\ii)^{k}}{2^{k} (k+1)!}(\phi: {\sf J})^{k}(V)_\mu
 = \sum_{k=0}^\infty \frac{1}{(k+1)!}\underbrace{\left(\phi_{\mu\nu_1} 
 \phi^{\nu_1\nu_2}\ldots \phi_{\nu_{k-1}\nu_{k}}\right)}_\text{k times} V^{\nu_k} .
$$
The eq.~\eqref{intermed2} can be read as the factorization of the exponential
of a linear combinations of generators of the Poincar\'e group. For this reason, 
it must be derivable also by using the known formulae of the factorization of the 
exponential of the sum of matrices $\exp[A+B]$ in terms of exponentials of commutators 
of $A$ and $B$. Indeed, it can be shown, by 
using the commutation relations of $\wP$ and $\wJ$, that one precisely gets the 
eq.~\eqref{intermed2} for {\em any} vector $V$ and tensor $\phi$, either real or 
complex. Hence, the formula \eqref{intermed2} can be applied to factorize the density 
operator \eqref{general}, neglecting $\h Q$ for the moment, by setting $\phi = \ii \varpi$:
\be\label{dopfact}
\wrho = \frac{1}{Z} \exp[- b \cdot \widehat P + \varpi : \wJ/2]
 = \frac{1}{Z} \exp[-\tilde b(\varpi) \cdot \widehat P] \exp[\varpi : \wJ/2]
\ee
with:
\be\label{btilde}
  \tilde b(\varpi)_\mu \equiv \sum_{k=0}^\infty \frac{1}{2^{k} (k+1)!}(\varpi: {\sf J})^{k}(b)_\mu
 = \sum_{k=0}^\infty \frac{\ii^k}{(k+1)!}\underbrace{\left(\varpi_{\mu\nu_1} 
 \varpi^{\nu_1\nu_2}\ldots \varpi_{\nu_{k-1}\nu_{k}}\right)}_\text{k times} b^{\nu_k}.
\ee
It is important to point out that $\tilde b(\varpi)$ is in general a complex
vector, however, according to eq.~\eqref{btilde}, it is {\em real} when $\varpi$ is
imaginary, that is when we deal with actual Lorentz transformations. 

\subsection{Iterative solution with imaginary \texorpdfstring{$\varpi$}{vorticity}}

We can now take advantage of this factorized form to calculate the mean value
\eqref{mainmean}. The idea is to seek the solution for pure imaginary $\varpi$ first
and then to analytically continue it to real $\varpi$, the physical case. The reason 
is that with imaginary $\varpi = - \ii \phi$ one deals with an actual Lorentz transformation, 
for which commutation relations with creation and annihilation operators are known.
Specifically, a Lorentz transformation operator $\widehat{\sf \Lambda}=\exp[-\ii\phi:\widehat J/2]$
acts as\footnote{Throughout this work, ${\sf \Lambda}p$ is a shorthand for ${\sf \Lambda}(p)$}:
\be\label{comm1}
 \widehat{\sf \Lambda}\, \wad{p} \widehat{\sf \Lambda}^{-1} = \wad{{\sf \Lambda}p} .
\ee
For the translation-like operators, the commutation relations can be calculated
regardless of whether the vector $\tilde b$ is real or complex:
\be\label{comm2}
 \exp[-\tilde b \cdot \widehat{P}]\, \wad{p} \exp[\tilde b \cdot \widehat{P}]
  = \e^{-\tilde b \cdot p} \, \wad{p}.
\ee

Let us then write the analytic continuation of the relation \eqref{dopfact} to 
$\varpi = -\ii \phi$, so:
$$
\wrho = \frac{1}{Z} \exp[- b \cdot \widehat P -\ii \phi : \widehat J/2]
 = \frac{1}{Z} \exp[-\tilde b (-\ii \phi) \cdot \widehat P] \exp[-\ii \phi : \widehat J/2]
 \equiv  \frac{1}{Z}\exp[-\tilde b(-\ii \phi) \cdot \widehat P] \, \widehat{\sf \Lambda}.
$$
By using the relations \eqref{comm1} and \eqref{comm2}, the analytic continuation 
of \eqref{mainmean} can be worked out as follows:
\begin{align*}
 & \Tr \left( \wrho \; \wad{p} \, \wa{p^\prime} \right) = \frac{1}{Z} \Tr \left(
 \exp[-\tilde b \cdot \widehat P] \, \widehat{\sf \Lambda} \, \wad{p} \, \wa{p^\prime}
 \right) = \frac{1}{Z} \Tr \left( \exp[-\tilde b \cdot \widehat P] \, \wad{{\sf \Lambda}p}
 \widehat{\sf \Lambda} \, \wa{p^\prime} \right) \\
 & = \frac{1}{Z} \e^{-\tilde b \cdot {\sf \Lambda}(p)}\, 
 \Tr \left( \wad{{\sf \Lambda}p} \, \exp[-\tilde b \cdot \widehat P] \,
 \widehat{\sf \Lambda} \, \wa{p^\prime} \right) = \e^{-\tilde b \cdot {\sf \Lambda}(p)} \,
 \Tr \left( \wa{p^\prime} \, \wad{{\sf\Lambda}p} \, \wrho \right) =  
 \e^{-\tilde b \cdot {\sf \Lambda}(p)} \Tr \left( \wrho \, \wa{p^\prime} \, \wad{{\sf\Lambda}p}
 \right) ,
\end{align*}
where, in the last equalities, we have used the ciclicity of trace. The above 
derivation can be written by using the shorthand of $\langle \; \rangle$ replacing the
trace with the density operator:
\begin{equation*}
 \media{\wad{p} \, \wa{p^\prime}} = \e^{-\tilde b \cdot {\sf\Lambda}(p)} 
 \media{\wa{p^\prime} \, \wad{{\sf\Lambda}p}}.
\end{equation*}
We can now use the known commutation relation between creation and annihilation operators
$$
  [\wa{p} , \wad{p^\prime}] = \delta^3 ({\bf p}-{\bf p}^\prime) 2\varepsilon^\prime,
$$
where $\varepsilon = \sqrt{{\bf p}^2+m^2}$, to obtain:
\be\label{basic}
  \media{\wad{p} \, \wa{p^\prime}} = \e^{-\tilde b \cdot {\sf \Lambda}p}
   \media{\wad{{\sf\Lambda}p} \wa{p^\prime}} + 
   \delta^3({\sf\Lambda}{\bf p} - {\bf p}^\prime) 2 \varepsilon^\prime 
   \e^{-\tilde b \cdot p^\prime},
\ee  
where ${\sf \Lambda}{\bf p}$ stand for the space part of the four-vector ${\sf\Lambda}p$. This
is the basic equation we have to solve.

First, we observe that if $\sf \Lambda = \sf I$, we get back to the familiar thermal field
theory and the \eqref{basic} can be solved algebraically leading to the well known Bose-Einstein
distribution function, multiplied by $2\varepsilon$ in view of our adopted covariant 
commutation relations. In the more general case, we will obtain a consistent solution 
of \eqref{basic} by an iterative method. The idea is to approximate the left hand side
with the second term of the right hand side with the Dirac delta, and correct the expression
obtained by inserting the first term calculated with the previous approximation.
Let us see how this is accomplished in formulae. We start with:
$$
 \media{\wad{p} \, \wa{p^\prime}} \simeq \delta^3({\sf\Lambda}{\bf p} - {\bf p}^\prime) 
 2 \varepsilon^\prime \e^{-\tilde b \cdot p^\prime}.
$$
This approximated solution implies:
$$
 \media{\wad{{\sf \Lambda}p} \, \wa{p^\prime}} \simeq \delta^3({\sf\Lambda}^2{\bf p} 
 - {\bf p}^\prime) 2 \varepsilon^\prime \e^{-\tilde b \cdot p^\prime} .
$$
Now we can plug the above expression again in the \eqref{basic} and get, after 
multiplying both sides by $\e^{\tilde b \cdot p^\prime}$:
\be\label{iter2}
\e^{\tilde b \cdot p^\prime} \media{\wad{p} \, \wa{p^\prime}} \simeq 
 \delta^3({\sf\Lambda}^2{\bf p} - {\bf p}^\prime) \e^{-\tilde b \cdot {\sf \Lambda}(p)} 
 2 \varepsilon^\prime + \delta^3({\sf\Lambda}{\bf p} - {\bf p}^\prime) 2 \varepsilon^\prime.
\ee  
From eq.~\eqref{iter2} we can now calculate an updated approximated expression of 
$\media{\wad{{\sf \Lambda}(p)} \, \wa{p^\prime}}$:  
$$
 \e^{\tilde b \cdot p^\prime} \media{\wad{{\sf \Lambda}(p)} \, \wa{p^\prime}} 
  \simeq \delta^3({\sf\Lambda}^3{\bf p} - {\bf p}^\prime) \e^{-\tilde b \cdot {\sf  \Lambda}^2(p)} 
  2 \varepsilon^\prime + \delta^3({\sf\Lambda}^2{\bf p} - {\bf p}^\prime) 2 \varepsilon^\prime
$$
and plugging it again in eq.~\eqref{basic}; we obtain:
$$
\e^{\tilde b \cdot p^\prime} \media{\wad{p} \, \wa{p^\prime}} \simeq 
\left[ \delta^3({\sf\Lambda}^3{\bf p} - {\bf p}^\prime) \e^{-\tilde b \cdot {\sf \Lambda}^2(p)
- \tilde b \cdot {\sf \Lambda}p} + \delta^3({\sf\Lambda}^2{\bf p} - {\bf p}^\prime) 
\e^{- \tilde b \cdot {\sf \Lambda}p} + \delta^3({\sf\Lambda}{\bf p} - {\bf p}^\prime) \right]
2 \varepsilon^\prime.
$$
Iterating, we eventually obtain the solution of \eqref{basic} in the form of a series:
\be\label{iterlast}
 \media{\wad{p} \, \wa{p^\prime}} = 2 \varepsilon^\prime \sum_{n=1}^\infty
 \delta^3({\sf\Lambda}^n{\bf p} - {\bf p}^\prime) \exp\left[ -\sum_{k=1}^{n}  
 \tilde b \cdot {\sf \Lambda}^k p \right] .
\ee
The expression \eqref{iterlast} can be further transformed by using the orthogonality of
$\sf \Lambda$:
$$
  {\sf \Lambda} x \cdot {\sf \Lambda} y = x \cdot y.
$$  
Note that, the orthogonality relation holds for complex $\varpi$ and complex vectors $x,y$, so this
equality would not be affected by choosing $\varpi$ real or imaginary. We can then rewrite 
eq.~\eqref{iterlast} as:
\be\label{final}
 \media{\wad{p} \, \wa{p^\prime}} =2 \varepsilon^\prime \sum_{n=1}^\infty
 \delta^3({\sf\Lambda}^n{\bf p} - {\bf p}^\prime) \exp \left[ -\sum_{k=1}^{n}  
 {\sf \Lambda}^{-k} \tilde b \cdot p \right].
\ee
It is possible to find alternative forms of the \eqref{final} without the appearance of Lorentz 
transformations in the exponent. Indeed, it can be shown that (see Appendix \ref{recurrence}), 
for generally complex $\varpi$ such that ${\sf \Lambda} = \exp [\varpi : {\sf J}/2]$:
\be\label{recurr}
  {\sf \Lambda}^{-k} \tilde b (\varpi) = (1-k) \tilde b \left( (1-k)\varpi \right) + 
  k \tilde b\left(-k \varpi \right),
\ee
implying:
\be\label{recurr2}
  \sum_{k=1}^n {\sf \Lambda}^{-k} \tilde b (\varpi) = n \tilde b(-n\varpi) .
\ee
By substituting the \eqref{recurr2} into the \eqref{final}, we obtain its alternative 
form:
\be\label{final2}
 \media{\wad{p} \, \wa{p^\prime}} = 2 \varepsilon^\prime \sum_{n=1}^\infty
 \delta^3({\sf\Lambda}^n{\bf p} - {\bf p}^\prime) \exp[ - n \,\tilde b (-n \varpi)\cdot p] .
\ee 

Some comments are now in order. First of all, it can be checked that indeed the found
series in either form \eqref{final} or \eqref{final2} are actual solution of the equation
\eqref{basic} by direct substitution. Besides, it can be readily realized that they are
an extension of the Bose distribution function by taking the limit $\varpi \to 0$, 
which implies ${\sf \Lambda} \to {\sf I}$. Since, from \eqref{btilde} $\tilde b(0) = b$
one has, from \eqref{final2}: 
$$
 \media{\wad{p} \, \wa{p^\prime}} = 2 \varepsilon^\prime \delta^3({\bf p} - {\bf p}^\prime)
  \sum_{n=1}^\infty \exp[-n \,b \cdot p] = 2 \varepsilon^\prime \frac{1}{\e^{b \cdot p} - 1}
  \delta^3({\bf p} - {\bf p}^\prime).
$$

An important point is that, because of the Dirac deltas with argument ${\sf \Lambda}{\bf p}$, 
the forms \eqref{final},\eqref{final2} for finite $\varpi$ make sense only if $\varpi$ is 
imaginary, that is only if ${\sf \Lambda}$ is an actual Lorentz transformation.
Therefore, it is possible to make analytic continuations only {\em after} the integration 
in the momentum $p'$ is done. Once the integration is carried out, the convergence of 
the analytically continued series is not trivial (see later discussion). We just remark that 
if $\tilde b(\varpi)$ is time-like and future oriented, each exponent in the \eqref{basic} is 
negative, what is more apparent in the form \eqref{iterlast}. 

To conclude, we observe that, should the density operator include a chemical potential 
term coupled to a conserved charge like in eq.~\eqref{general}, the iterative derivation 
can be repeated and one has one more exponential factor:
\be\label{final3}
  \media{\wad{p} \, \wa{p^\prime}} = 2 \varepsilon^\prime \sum_{n=1}^\infty
  \delta^3({\sf\Lambda}^n{\bf p} - {\bf p}^\prime) \exp[ - n\, \tilde b (-n \varpi)\cdot p + 
  n\, \zeta].
\ee 

An important question is whether the \eqref{final}, or one of its equivalent forms
\eqref{final2} \eqref{final3}, is the unique solution of equation \eqref{basic} or if there
are possibly more solutions. In general, it can be readily shown that any solution $S(p,p')$
of the \eqref{basic} can be written as:
$$
 S(p,p') = H(p,p') + S_0(p,p')
$$
where $S_0(p,p')$ is a particular solution of the \eqref{basic}, such as \eqref{final3}, 
and $H(p,p')$ is the general solution of the associated homogeneous equation:
\be\label{homog}
  \media{\wad{p} \, \wa{p^\prime}} = \e^{-\tilde b \cdot {\sf \Lambda}p}
   \media{\wad{{\sf\Lambda}p}\, \wa{p^\prime}}.
\ee  
We will see that the existence of non-trivial solutions of the homogeneous equation 
\eqref{homog} depends on the Lorentz transformation $\sf \Lambda$; if $\sf \Lambda$ is 
a rotation, we will show in Section \ref{rotation} that there are no non-trivial solutions; 
conversely, if $\sf \Lambda$ is a pure boost, there are non-trivial solutions, as shown
in Section \ref{sec:accel}. 

Nevertheless, it is a general feature that solutions of the homogeneous equation \eqref{homog} 
are non-analytic in thermal vorticity around zero, that is for $\phi= \ii \varpi = 0$, 
as shown in Section \ref{math}. Therefore, the particular solution found by iteration \eqref{basic} 
may, in principle, contain non-analytic contributions in $\varpi$ around zero which would obviously 
hinder a process of analytic continuation from imaginary to real $\varpi$. If we could single out 
an analytic part around ${\sf \Lambda} = {\sf I}$ of the particular solution \eqref{basic}, the function 
would be unique and its analytic continuation possible by construction. In principle, 
there might be cases where no analytic solutions exist because $\varpi=0$ is itself a singular point 
in the thermodynamics. However, to our knowledge, no instances of this phenomenon are known, hence
the physical solution can be determined by extracting the analytic part. This method of {\em analytic 
distillation} is discussed in Section \ref{math}.

\section{The covariant Wigner function and the phase space distribution function}
\label{phasespace}

The expectation value of the quadratic combination \eqref{mainmean} is all we need to 
calculate every field-related quantity, like the stress-energy tensor or any other conserved current. 
However, it is convenient to make use of a very useful concept, the covariant Wigner function 
\cite{degroot}:
\be\label{wigner}
 W(x,k) = \frac{2}{(2\pi)^4} \int \di^4 y \; \media{:\wpsi^\dagger(x+y/2) \wpsi(x-y/2):} 
 \e^{-\ii y \cdot k},
\ee
where  ``$:\,:$'' stands for the normal ordering of creation and annihilation operators and a
density operator is understood in the mean value. Note that the covariant Wigner function
is real, but it is not positive definite. The covariant Wigner function 
\eqref{wigner} allows to express the mean particle current as a four-dimensional integral:
\be\label{wigcurr}
 j^\mu(x) =  \ii \media{: \wpsi^\dagger(x) \codevmu \wpsi(x):} = \int \di^4 k \; k^\mu W(x,k).
\ee
Plugging the free scalar field expansion: 
\be\label{field}
 \wpsi(x) = \frac{1}{(2\pi)^{3/2}} \int \frac{\di^3 \p}{2\varepsilon} \left[
 \e^{-\ii p \cdot x} \wa{p} + \e^{\ii p \cdot x} \wbd{p}\right]
\ee
in the eq.~\eqref{wigner} we get:
\begin{align}\label{wigner2}
 W(x,k) &= \frac{2}{(2\pi)^{7}} \int \frac{\di^3 \p}{2\varepsilon} \frac{\di^3 \p^\prime}
 {2\varepsilon^\prime}\int \di^4 y \left[
 \e^{-\ii y \cdot (k - (p+p^\prime)/2)} \e^{\ii (p-p^\prime)\cdot x}
 \media{\wad{p}\wa{p^\prime}} + \e^{-\ii y \cdot (k + (p+p^\prime)/2)} \e^{-\ii (p-p^\prime)\cdot x}
 \media{\wbd{p^\prime}\wb{p}} \nonumber \right.\\
 &\left.+ \e^{-\ii y \cdot (k - (p-p^\prime)/2)} \e^{\ii (p + p^\prime)\cdot x}
 \media{\wad{p}\wbd{p^\prime}} + \e^{-\ii y \cdot (k + (p - p^\prime)/2)} \e^{-\ii (p+p^\prime)\cdot x}
 \media{\wb{p}\wa{p^\prime}}\right] \nonumber \\
 &= \frac{2}{(2\pi)^{7}} \int \frac{\di^3 \p}{2\varepsilon} \frac{\di^3 \p^\prime}
 {2\varepsilon^\prime}\int \di^4 y \left[
 \e^{-\ii y \cdot (k - (p+p^\prime)/2)} \e^{\ii (p-p^\prime)\cdot x}
 \media{\wad{p}\wa{p^\prime}} + \e^{-\ii y \cdot (k + (p+p^\prime)/2)} \e^{\ii (p-p^\prime)\cdot x}
 \media{\wbd{p}\wb{p^\prime}} \right.\nonumber \\
 &\left.+ \e^{-\ii y \cdot (k - (p-p^\prime)/2)} \e^{\ii (p + p^\prime)\cdot x}
 \media{\wad{p}\wbd{p^\prime}} + \e^{-\ii y \cdot (k - (p - p^\prime)/2)} \e^{-\ii (p+p^\prime)\cdot x}
 \media{\wb{p^\prime}\wa{p}} \right]\nonumber \\
 &= \frac{2}{(2\pi)^{3}} \int \frac{\di^3 \p}{2\varepsilon} \frac{\di^3 \p^\prime}{2\varepsilon^\prime}
\left[\e^{\ii (p-p^\prime)\cdot x} \left( \delta^4(k - (p+p^\prime)/2) \media{\wad{p}\wa{p^\prime}}
 + \delta^4(k + (p+p^\prime)/2) \media{\wbd{p}\wb{p^\prime}} \right) \right.\nonumber \\
 &\left.+ \delta^4 (k - (p - p^\prime)/2) \left( \e^{\ii (p + p^\prime)\cdot x} 
\media{\wad{p}\wbd{p^\prime}} +  \e^{- \ii (p + p^\prime)\cdot x} \media{\wb{p^\prime}\wa{p}} \right)
\right].
\end{align}
In the above equalities, we have taken advantage of the symmetric integration in the variables $p,\,p^\prime$.
The expression~\eqref{wigner2} makes it apparent that the variable $k$ of the Wigner function $W(x,k)$ 
is {\em not} on-shell, i.e. $k^2 \ne m^2$ even in the free case. This makes the definition of 
a particle distribution function $f(x,p)$ \`a la Boltzmann not straighforward in a quantum relativistic 
framework. From equation~\eqref{wigner2}, we can infer that $W$ is made up of three terms which
can be distinguished for the characteristic of $k$. For future time-like $k = (p+p^\prime)/2$, 
only the first term involving particles is retained; for past time-like $k= - (p+p^\prime)/2$, 
only the second term involving antiparticles; finally, for space-like $k = (p-p^\prime)/2$ 
the last term with the mean values of two creation/annihilation operators is retained. 
In symbols: 
\be\label{wdecomp}
    W(x,k ) = W(x,k)\theta(k^2)\theta(k^0) + W(x,k) \theta(k^2) \theta(-k^0) +
    W(x,k) \theta(-k^2) \equiv  W_+(x,k) + W_-(x,k) + W_S(x,k).
\ee
Multiplying the previous expression by $\theta(k^0)\theta(k^2)$ selects the particle contribution 
to the covariant Wigner function and by $\theta(-k^0)\theta(k^2)$ the anti-particle one. It is thus 
possible to define the particle covariant Wigner function $W_+(x,k)$ as well as the antiparticle 
counterpart $W_-(x,k)$. The particle contribution to the current can be obtained from eq.~\eqref{wigcurr} 
by using the \eqref{wigner2}:
\begin{align}\label{current}
 j_+^\mu(x) &= \frac{1}{(2\pi)^{3}} \int \frac{\di^3 \p}{\varepsilon} 
  \frac{\di^3 \p^\prime}{2\varepsilon^\prime} \;  \frac{(p + p^\prime)^\mu}{2} \, 
   \e^{\ii (p-p^\prime)\cdot x} \media{\wad{p}\wa{p^\prime}} \nonumber \\
 &= \int \frac{\di^3 \p}{\varepsilon} p^\mu {\rm Re} \left(\frac{1}{(2\pi)^3} 
  \int \frac{\di^3 \p^\prime}{2\varepsilon^\prime} \; \e^{\ii (p-p^\prime)\cdot x} 
  \media{\wad{p}\wa{p^\prime}} \right).
\end{align}
To obtain the last expression, we have taken advantage of the hermiticity of the density operator, 
implying:
$$
  \media{\wad{p}\wa{p^\prime}} = \media{\wad{p^\prime}\wa{p}}^*
$$
which makes it possible to swap the integration variables $p$ and $p^\prime$. The formula \eqref{current} 
identifies the particle distribution function or phase space density $f(x,p)$ as the real part of a 
{\em complex distribution function} $f_c(x,p)$:
\be\label{phspace}
 f(x,p) = {\rm Re} f_c(x,p),
\ee
where
\be\label{phspacec}
 f_c(x,p) = \frac{1}{(2\pi)^3} \int \frac{\di^3 \p^\prime}{2\varepsilon^\prime} 
   \; \e^{\ii (p-p^\prime)\cdot x} \media{\wad{p}\wa{p^\prime}}.
\ee
For antiparticles the distribution function $\bar f(x,p)$ is obtained from \eqref{phspace} replacing $\media{\wad{p}\wa{p^\prime}}$ with $\media{\wbd{p}\wb{p^\prime}}$. Also note that with the density operator 
\eqref{general} all terms involving pairs of creation and annihilation operators in \eqref{wigner2} vanish 
according to the \eqref{nulli}, so that the current does not involve any space-like contribution from $W_S(x,k)$ and
one is left with:
$$
 j^\mu(x) =  \int \di^4 k \; k^\mu W(x,k) = {\rm Re}
 \int \frac{\di^3 \p}{2\varepsilon} p^\mu \left[ f_c(x,p) - {\bar f}_c(x,p) \right]  ,
$$
which also shows that the current can be written as an integral of on-shell four-momenta when using the 
phase space densities and off-shell, equation~\eqref{wigcurr}, when using the Wigner function.

We can now turn to the calculation of the mean value of the stress-energy tensor. From 
the Lagrangian:
$$
 {\cal L} = \partial_\mu \wpsi^\dagger \partial^\mu \wpsi - m^2 \wpsi^\dagger \wpsi
$$
one can obtain the so-called {\em canonical} stress-energy tensor operator:
\be\label{canonical}
  \wT_C^{\mu\nu} = \partial^\mu \wpsi^\dagger \partial^\nu \wpsi + 
   \partial^\nu \wpsi^\dagger \partial^\mu \wpsi - g^{\mu\nu} {\cal L}
\ee
and so determine its normal ordered mean value $\langle : \wT_C^{\mu\nu} : \rangle$. 
For this purpose, we again make use of the covariant Wigner function and the known expression 
\cite{degroot}:
\be\label{degroot1}
  -\frac{1}{2} \langle : \wpsi^\dagger \codevmu \codevnu \wpsi : \rangle = 
  \int \di^4 k \; k^\mu k^\nu W(x,k).
\ee  
Now, it can be shown by using the equations of motion of the free scalar field, that:
$$
  \wT_C^{\mu\nu} = -\frac{1}{2} \wpsi^\dagger \codevmu \codevnu \wpsi + 
  \frac{1}{2} \partial^\mu \partial^\nu (\wpsi^\dagger \wpsi) - \frac{1}{2} g^{\mu\nu} 
  \Box (\wpsi^\dagger \wpsi).
$$
Since, from the covariant Wigner function definition \eqref{wigner}:
\be\label{degroot2}
   \langle : \wpsi^\dagger \wpsi : \rangle = \frac{1}{2} \int \di^4 k \; W(x,k),
\ee   
we finally obtain, by using \eqref{degroot1} and \eqref{degroot2}:
\be\label{cantens}
 \langle : \wT_C^{\mu\nu}: \rangle = \int \di^4 k \; k^\mu k^\nu W(x,k)
  + \frac{1}{4} \left( \partial^\mu \partial^\nu - g^{\mu\nu} \Box \right)
   \int \di^4 k \; W(x,k).
\ee
We note in passing that, since $\wT^{\mu\nu}$ is a quadratic operator of creation and 
annihilation operators of momentum eigenstates, just like the Wigner operator, its normally ordered 
mean value corresponds to the subtraction of the vacuum expectation value from the plain mean 
value, that is:
$$
 \langle : \wT_C^{\mu\nu} : \rangle =  \langle  \wT_C^{\mu\nu} \rangle -
  \bra{0}  \wT_C^{\mu\nu} \ket{0}.
$$

In order to derive the relation between the stress-energy tensor and the phase space 
distribution function, we first write down the derivative of $f_c$ in the eq.~\eqref{phspacec}:
$$
 \partial_\nu f_c (x,p) = \frac{1}{(2\pi)^3} \int \frac{\di^3 \p^\prime}{2\varepsilon^\prime} 
   \; \ii (p-p^\prime)_\nu \e^{\ii (p-p^\prime)\cdot x} \media{\wad{p}\wa{p^\prime}},
$$
whence:
\be\label{gradf}
  \frac{1}{(2\pi)^3} \int \frac{\di^3 \p^\prime}{2\varepsilon^\prime} p^\prime_\nu 
  \e^{\ii (p-p^\prime)\cdot x} \media{\wad{p}\wa{p^\prime}} = \ii \partial_\nu f_c + p_\nu f_c \,.
\ee
Now we can plug the \eqref{wigner2} in \eqref{cantens} and by the same method used for
the current, we can express the stress-energy tensor as a function 
of the phase space distribution function:
\be\label{cantens2}
 \langle : \wT_C^{\mu\nu}: \rangle = {\rm Re} \left[ 
  \int \frac{\di^3 p}{\varepsilon} \; \left( p^\mu p^\nu + \frac{1}{4} 
  \left( \ii p^\mu \partial^\nu + \ii p^\nu \partial^\mu \right) +
  \frac{1}{4} \left( \partial^\mu \partial^\nu - g^{\mu\nu} \Box \right) \right) 
  (f_c(x,p) + \bar f_c(x,p)) \right].
\ee
The above expression makes it apparent that the gradient terms are quantum correction to 
the ``classical'' expression of the stress-energy tensor in kinetic theory in that they
require - after restoring the natural constants - $\hbar,\hbar^2$ factors to have the same 
dimension. Also, the phase space distribution functions may themselves have quantum 
corrections depending on $\hbar$. It should also be emphasized that the relation between 
stress-energy tensor and phase space distribution, unlike sometimes believed, depends on the 
particular stress-energy tensor operator. This relation indeed is non-invariant under 
pseudo-gauge transformations except at homogeneous thermodynamic equilibrium \cite{Becattini:2011ev}.

The equation \eqref{final} or its equivalent forms~\eqref{final2},\eqref{final3} can now be used 
to calculate the complex phase space distribution function \eqref{phspacec} for the 
general global equilibrium. It is important to keep in mind that the expectation values
were obtained for imaginary $\varpi$ and that an analytic continuation to real $\varpi$ 
must be eventually carried out. By using the eq.~\eqref{final2}, it turns out to be:
\begin{align}\label{phspacec3}
 f_c(x,p) &=  \frac{1}{(2\pi)^3} 
 \sum_{n=1}^\infty \exp[\ii ({\sf I}-{\sf \Lambda}^n)p \cdot x] 
 \exp[ - n\, \tilde b (-n \varpi) \cdot p + n\,\zeta] \nonumber \\
 &=  \frac{1}{(2\pi)^3} \sum_{n=1}^\infty \exp[({\sf I}-{\sf \Lambda}^{-n}) \ii x \cdot p ] 
 \exp[ - n\, \tilde b (-n \varpi) \cdot p + n\,\zeta],
\end{align} 
where we have used the orthogonality of $\sf \Lambda$ and its linearity. Since:
\be\label{useful1}
 {\sf I} - {\sf \Lambda}^{-n} = \sum_{k=0}^{n-1} {\sf \Lambda}^{-k} ({\sf I} - {\sf \Lambda}^{-1})
\ee
and:
\be\label{useful2}
   \left( \frac{\varpi: {\sf J}}{2} \right) x = \ii \varpi \cdot x ,
\ee
we have:
\be\label{useful3}
 ({\sf I} - {\sf \Lambda}^{-1})\ii x = -\sum_{k=1}^\infty \frac{1}{k!} \left( 
  \frac{-\varpi:{\sf J}}{2}\right)^k \ii x = - \sum_{k=1}^\infty \frac{1}{k!} \left( 
  \frac{- \varpi:{\sf J}}{2}\right)^{k-1} \varpi \cdot x = - \sum_{k=0}^\infty \frac{1}{(k+1)!} \left( 
  \frac{-\varpi:{\sf J}}{2}\right)^{k} (\varpi \cdot x) = - \tilde c (-\varpi),
\ee 
where we have defined $c \equiv \varpi \cdot x$ and used the eq.~\eqref{btilde} as a
definition of tilde-transformed vector. Therefore, by using \eqref{useful1}, \eqref{useful3}
and \eqref{recurr2}, which hold for any tilde-transformed vector, we obtain:
$$
 ({\sf I}-{\sf \Lambda}^{-n}) \ii x \cdot p = - \sum_{k=0}^{n-1} {\sf \Lambda}^{-k} 
 \tilde c (-\varpi) \cdot p = - n \tilde c (-n\varpi) \cdot p.
$$
We can substitute this result into the \eqref{phspacec3}:
\begin{equation*}
 f_c(x,p) =  \frac{1}{(2\pi)^3} 
 \sum_{n=1}^\infty \exp[ - n (\tilde b(-n\varpi) + \tilde c(-n \varpi)) \cdot p 
 + n\,\zeta ] .
\end{equation*}
Now note that, by definition of $\tilde b$ and $\tilde c$:
\be\label{tildebeta}
\tilde b(\varpi) + \tilde c(\varpi) = \sum_{k=0}^\infty \frac{1}{(k+1)!} 
 \left(\frac{\varpi:{\sf J}}{2}\right)^{k} (b + \varpi \cdot x) = 
 \sum_{k=0}^\infty \frac{1}{(k+1)!} \left(\frac{\varpi:{\sf J}}{2}\right)^{k} \beta
 = \tilde\beta (\varpi),
\ee
where $\beta$ is the four-temperature vector of global thermodynamic equilibrium
in eq. \eqref{fourtemp}. Like for the $\tilde b$ in eq.~\eqref{btilde}, the vector $\tilde\beta$
can be written as:
\be\label{tildebetaf}
   {\tilde \beta}_\mu = \sum_{k=0}^\infty \frac{1}{(k+1)!} \left(\frac{\varpi:{\sf J}}{2}\right)^{k} 
  (\beta)_\mu
   = \sum_{k=0}^\infty \frac{\ii^k}{(k+1)!}\underbrace{\left(\varpi_{\mu\nu_1} 
 \varpi^{\nu_1\nu_2}\ldots \varpi_{\nu_{k-1}\nu_{k}}\right)}_\text{k times} \beta^{\nu_k}
\ee
Another very useful expression is:
\be\label{tildebeta2}
  \tilde \beta(\varpi) = \tilde b (\varpi) + \ii ({\sf I} - \e^{\varpi: {\sf J}/2}) x
\ee
which is a consequence of \eqref{useful2} and \eqref{tildebeta}.

We can finally write the complex phase space distribution function as:
\be\label{phspacec4}
\boxed{
 f_c(x,p) = \frac{1}{(2\pi)^3} 
 \sum_{n=1}^\infty \exp[ - n\, \tilde \beta (-n \varpi) \cdot p + n\,\zeta ].
}
\ee
For antiparticles, the $\bar f_c(x,p)$ is obtained by replacing $\zeta \to -\zeta$.
We are now going to study the above series and highlight some of its main features.

\section{Properties of the distribution function}
\label{properties}

From the function \eqref{phspacec4} we can calculate most quantities of interest, once 
analytic continuation to real $\varpi$ is made. 

We first observe that in the limit $\varpi \to 0$ the series is real and it boils down to 
the Bose-Einstein distribution function:
$$
  f_c(x,p) = \frac{1}{(2\pi)^3} \sum_{n=1}^\infty \exp (- n b \cdot p + n \zeta)
   = \frac{1}{(2\pi)^3} \frac{1}{\e^{b\cdot p - \zeta} - 1},
$$
where the last equality applies if $\zeta$ is lower than the mass so that the series
converges. This suggests that the series expresses the familiar quantum statistics 
expansion and its first term is the Boltzmann limit.

Otherwise, if $\varpi \ne 0$, the \eqref{phspacec4} is a series of analytic functions of 
the various $\varpi^{\mu\nu}$ taken as complex variables and if the series uniformly converges 
in some region, it defines an analytic function therein. 
In fact, in the physical case with $\varpi$ real, the series in eq.~\eqref{phspacec4}
is not always convergent, as we will see, whether $\tilde b(\varpi)$ is time-like or not. 
While this does not prevent an analytic continuation of the phase space distribution function,
it requires a careful analysis to obtain it. Particularly, it turns out that the properties 
of the series are different according to which components appear in the thermal vorticity 
tensor $\varpi$ and a separate treatment is needed. 

In the physical case, it is worth pointing out that ${\rm Re} \tilde\beta$ is itself a 
Killing vector, just like $\beta$. Looking at the equation in \eqref{tildebetaf} 
we obtain:
$$
 \partial_\nu \tilde \beta_\mu = \sum_{k=0}^\infty \frac{\ii^k}{(k+1)!}\underbrace{\left(\varpi_{\mu\nu_1} 
 \varpi^{\nu_1\nu_2}\ldots \varpi_{\nu_{k-1}\nu_{k}}\right)}_\text{k times} \partial_\nu \beta^{\nu_k}
 =  - \sum_{k=0}^\infty \frac{\ii^k}{(k+1)!}\underbrace{\left(\varpi_{\mu\nu_1} 
 \varpi^{\nu_1\nu_2}\ldots \varpi_{\nu_{k-1}\nu_{k}}\varpi^{\nu_k}_{\, \nu} \right)}_\text{k+1 times} .
$$
Each term of the last series is antisymmetric in $\mu \leftrightarrow \nu$ if $k$ is even and
symmetric if $k$ is odd. Thus, because of $\ii^k$, the real part of the right hand
side series selects antisymmetric terms and $\partial_\nu {\rm Re} \tilde \beta_\mu$ 
turns out to be an antisymmetric tensor, which proves our statement. Conversely, the
imaginary part of $\partial_\nu \tilde\beta_\mu$ is a symmetric tensor. This observation
leads to the conclusion that $f(x,p)$ is {\rm not} a solution of the Boltzmann equation.
From the eq.~\eqref{phspacec4}, with $\tilde\beta_n \equiv \tilde\beta(-n \varpi)$:
\begin{align*}
 \partial_\mu f &= -\frac{1}{(2\pi)^3} {\rm Re}
 \sum_{n=1}^\infty n\, p^\nu \partial_\mu \tilde\beta_{n\,\nu} 
  \exp[ - n \tilde \beta_n \cdot p + n\zeta ]\\ &= -\frac{1}{(2\pi)^3} 
 \sum_{n=1}^\infty n\, p^\nu \partial_\mu ({\rm Re } \tilde\beta_{n \,\nu}) 
  {\rm Re}\, \e^{- n \tilde \beta_n \cdot p + n\zeta} - n\, p^\nu
  \partial_\mu ({\rm Im} \tilde\beta_{n \, \nu})
  {\rm Im} \,  \e^{- n \tilde \beta_n \cdot p + n\zeta} .
\end{align*}
If we now multiply by $p^\mu$ we get:
$$
 p^\mu \partial_\mu f = -\frac{1}{(2\pi)^3} 
 \sum_{n=1}^\infty n\, p^\nu p^\mu \partial_\mu ({\rm Re } \tilde\beta_{n \,\nu}) 
  {\rm Re}\, \e^{- n \tilde \beta_n \cdot p + n\zeta} - n\, p^\nu p^\mu
  \partial_\mu ({\rm Im} \tilde\beta_{n \, \nu})
  {\rm Im} \,  \e^{- n \tilde \beta_n \cdot p + n\zeta}.
$$
The first term on the right hand side vanishes as ${\rm Re } \tilde\beta_n$ is 
a Killing vector, while the second term in general is non vanishing.  

It is also important to point out that the form \eqref{phspacec4} is indeed a resummation 
of all possible quantum corrections at all orders in $\hbar$ of the Bose-Einstein distribution 
function. Looking at \eqref{tildebeta2} it can be realized that, restoring natural constants, 
$\beta$ is a classical vector taking into 
account the eqs.~\eqref{fourtemp} and \eqref{thvort}, whereas the $k$-th term in the series
is proportional to $\hbar^k$ to make $\varpi : {\sf J}$ adimensional. Hence, an expansion in 
$\varpi$ of the \eqref{phspacec4} with fixed $\beta$, such as (setting $\zeta=0$ for simplicity):
\begin{align*}
 f_c(x,p) &= \frac{1}{(2\pi \hbar)^3} \sum_{n=1}^\infty \exp \left[ - n \tilde \beta (-n \varpi) 
 \cdot p \right]
 = \frac{1}{(2\pi)^3} \sum_{n=1}^\infty \exp \left[ - n \sum_{k=0}^\infty \frac{1}{(k+1)!} 
  \left(\frac{- n \hbar \varpi:{\sf J}}{2}\right)^{k} \beta \cdot p \right] \\
  & = \frac{1}{(2\pi \hbar)^3} \sum_{n=1}^\infty \prod_{k=0}^\infty \exp \left[ (- n)^{k+1} \frac{1}{(k+1)!} 
  \left(\frac{- \hbar \varpi:{\sf J}}{2}\right)^{k} \beta \cdot p \right] \\
  & = \frac{1}{(2\pi \hbar)^3} \sum_{n=1}^\infty \exp [ -n \beta \cdot p] \prod_{k=1}^\infty 
   \sum_{l_k = 0}^\infty \frac{1}{l_k!} \left[ (- n)^{k+1} \frac{1}{(k+1)!} 
  \left(\frac{- \hbar \varpi:{\sf J}}{2}\right)^{k} \beta \cdot p \right]^{l_k},
\end{align*}
is in fact the full semi-classical expansion in $\hbar$, which has been purposely restored 
for the occasion; each power of $\varpi$ involves a corresponding factor $\hbar$ to the same 
power. The series above can be rearranged as an asymptotic power series in $\hbar$ or $\varpi$, 
which, as we will see in Sections \ref{sec:accel} and \ref{rotation}, is mostly divergent.
Notwithstanding, it could be used to work out the quantum corrections of, say, the stress-energy 
tensor, with the eq.~\eqref{cantens2}, at some fixed order in $\hbar$ or $\varpi$. It is worth
remarking the similarity of this situation with that of the general, non-equilibrium, gradient 
expansions in relativistic hydrodynamics \cite{Heller:2020uuy}, whose the above expression is,
in a sense, a special case (with vanishing shear tensor) because:
$$
  \varpi_{\mu\nu} = \frac{1}{2} (\partial_\nu \beta_\mu - \partial_\mu \beta_\nu)
$$
from eq.~\eqref{fourtemp}.

\section{Mathematical interlude: analytic distillation and series resummation}
\label{math}

The solution of the equation \eqref{basic} found by iteration, the eq.~\eqref{final3}, may not be
unique if the associated homogeneous equation \eqref{homog} has non-trivial solutions, as we 
discussed in Section \ref{method}.
However, it is possible to show that if \eqref{homog} has non-trivial solutions they are non-analytic 
about the identity, that is for $\phi = \ii \varpi = 0$, being ${\sf \Lambda}= \exp[- \ii \phi : {\sf J}]$.

To prove it, suppose a non-trivial analytic solution around $\phi=0$ exists, $F(p,p^\prime,\phi)$
(without loss of generality, we can assume that $\phi$ is a single-valued variable). We can then write:
$$
 F(p,p^\prime,\phi) = \sum_{n=0}^\infty \frac{1}{n!} \frac{\partial^n F(p,p^\prime,\phi)}{\partial \phi^n} 
 \Bigg|_{\phi=0} \phi^n = \sum_{n=0}^\infty F_n(p,p^\prime) \phi^n 
$$
and equate both sides of \eqref{homog}:
$$
 \sum_{n=0}^\infty F_n(p,p^\prime) \phi^n = \e^{-\tilde b(\phi) \cdot {\sf \Lambda}(\phi)p}
  F({\sf\Lambda}(\phi)p, p^\prime,\phi).
$$
We can now expand the right hand side around $\phi=0$ and equate each power of $\phi$ separately. 
For the constant term at $\phi=0$ we get:
$$
 F_0(p,p^\prime) = \e^{-b \cdot p} F(p,p^\prime,0) = \e^{-b \cdot p} F_0(p,p^\prime),
$$
whose solution is $F_0(p,p^\prime)=0$. At first order we have, taking into account that $F_0(p,p^\prime) = 0$:
\begin{align*}
 & F_1(p,p^\prime) = \e^{-b \cdot p} \frac{\partial}{\partial p^\mu} F(p,p^\prime,0)
 (-\ii {\sf J}^\mu_\nu p^\nu) + \e^{-b \cdot p} F_1(p,p^\prime) = 
 \e^{-b \cdot p} \frac{\partial}{\partial p^\mu} F_0(p,p^\prime)
 (-\ii {\sf J}^\mu_\nu p^\nu) + \e^{-b \cdot p} F_1(p,p^\prime) \\
 &= \e^{-b \cdot p} F_1(p,p^\prime),
\end{align*}
whence $F_1(p,p^\prime)=0$. Iterating, by taking derivatives of higher order, it can be shown 
that all functions $F_n(p,p^\prime)$ vanish, hence the only analytic solution of the equation 
\eqref{homog} around $\phi=0$ is 0. In Appendix \ref{solhom} a particular non-analytic solution of the 
\eqref{homog} is found.

It should be now quite clear that non-analytic terms in $\varpi=0$ could be hidden in the 
particular solution \eqref{final3} and all quantities derived therefrom. This makes analytic 
continuations quite problematic unless an unambiguous procedure to single out an analytic
part is pinpointed. We call this procedure {\em analytic distillation} and we define it as
follows:
\begin{definition} 
{\em Let $f(z)$ be a function on a domain $D$ of the complex plane and $z_0 \in \bar D$ a point 
where the function may not be analytic. Suppose that asymptotic\footnote{We denote asymptotic
equality with the symbol $\sim$.} power series of $f(z)$ in $z-z_0$ 
exist in subsets $D_i \subset D$ such that $\cup_i D_i = D$: 
$$
  f(z) \sim \sum_{n} a^{(i)}_n (z-z_0)^n 
$$
where $n$ can take integer negative values. If the series formed with the common coefficients 
in the various subsets restricted to $n \ge 0$ has a positive radius of convergence, the analytic 
function defined by this power series is called analytic distillate of $f(z)$ in $z_0$ and it 
is denoted by $\dist_{z_0} f(z)$.}
\end{definition}
It is worth dwelling upon this definition. First, note that asymptotic power series in a point 
of a function can be constructed \cite{erdelyi} with iterative limits:
$$
  a_n = \lim_{z \to z_0} \frac{ f(z) - \sum_{i=N_{min}}^{n-1} a_i (z-z_0)^i}{(z-z_0)^n}.
$$
However, these limits may depend upon the argument of the complex number $z$ (e.g. $\exp(-1/z)$
for $z=0$). Indeed, it is a well known fact that asymptotic expansions are different in different
angular sectors centered in $z_0$ - the well known Stokes phenomenon~\cite{erdelyi}. The above
definition prescribes to retain the part of the power series asymptotic expansion which is common
to the various sectors or subsets $D_i$ covering the domain where the function is originally defined: 
$$
\bar{a}_n\equiv \begin{cases}
        a^{(i)}_n & \text{if } a^{(i)}_n=a^{(j)}_n,\, \forall i,j\\
        0 & \text{otherwise}
    \end{cases}\qquad
  \dist_{z_0} f(z)\equiv \sum_{n} \bar{a}_n (z-z_0)^n;
$$
if there is no common part, the distillate is simply zero. The strong requirement in the definition 
is the convergence of the asymptotic series, which is not usually the case, but it will apply in our 
cases of interest, as we will see. Also, from the definition it turns out that, if the function is 
analytic in $z_0$, then $\dist_{z_0} f(z) = f(z)$. If, on the 
other hand, the function has an isolated pole in $z_0$, the distillation extracts the positive powers 
of the Laurent series.

A simple example of analytic distillation is that of a function which is the sum of an analytic
function and a non-analytic function in $z=0$, for instance:
$$
 f(z) = g(z) + \e^{-1/z}.
$$
The analytic distillation of $f$ in zero is
$$
 \dist_0 f(z) = g(z)
$$
because it can be readily shown that $\dist_0 \e^{-1/z} = 0$, as
its power series about zero in the domain ${\rm Re}(z) > 0$ is vanishing.

The calculation of an analytic distillate involves that of an asymptotic power series, which is
not always a simple task. When the function presents itself as a series of simple functions, however, 
there is an important result by D.~B.~Zagier \cite{zagier,Dorigoni:2020oon} which can be cast as follows:
\begin{theorem}\label{th1}
{\em Let $f: (0,+\infty) \to \mathbb{C}$ be a $C^\infty$ complex valued function on the positive real 
axis with the asymptotic power series in $x=0$ 
$$
  f(x) \sim \sum_{n=0}^\infty a_n x^n
$$  
and $\mathcal{O}(1/x^{1+\epsilon})$ for $x \to +\infty$ with $\epsilon > 0$. Then, the function defined 
by the series:
$$
   g(x) = \sum_{n=1}^\infty f(nx)
$$
has the asymptotic expansion for $x \to 0^+$:
$$
   g(x) \sim \frac{I_f}{x} + \sum_{n=0}^\infty a_n \zeta(-n) x^n ,
$$   
where 
$$
  I_f = \int_0^\infty \di x \; f(x)
$$
and $\zeta$ is the Riemann Zeta function. }
\end{theorem}
%
\begin{figure}[tbh]
	\centering
	\includegraphics{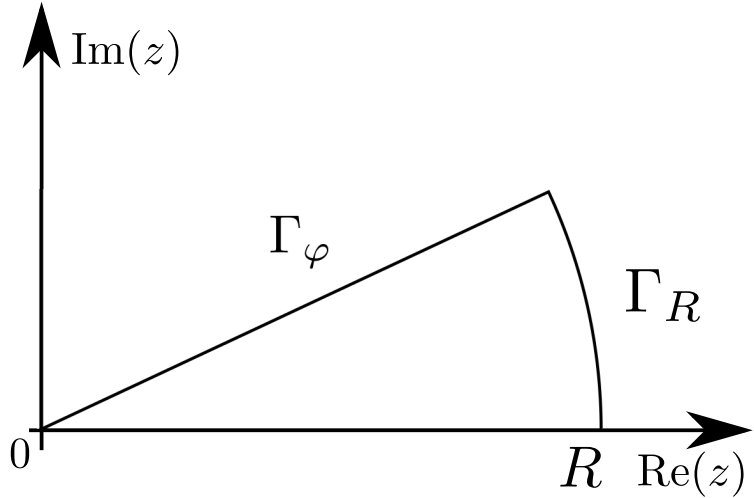}
	\caption{The path $\Gamma_\varphi$.}
	\label{fig:paths}
\end{figure}

This theorem can be extended to a complex function. Let $F(z)$ be a function of the complex 
variable $z$ with the asymptotic power series about $z=0$ in a domain of the complex plane 
including the real positive axis:
$$
  F(z) \sim \sum_{n=0} A_n z^n.
$$
Defining:
$$
  x = |z|  \implies z = x \, \e^{\ii \varphi} \, ,
$$
we can apply the previous theorem to the function $f(x)$ defined as:
$$
  f(x) \equiv F(x \, \e^{\ii \varphi}).
$$
We have, from the above definition, for the function $f(x)$: 
$$
a_n = A_n \e^{\ii n \varphi}
$$
and
$$
 I_f = \int_0^\infty \di x \; f(x) = \int_0^\infty \di x \; F(x \, \e^{\ii \varphi})
 = \e^{-\ii \varphi} \int_{\Gamma_\varphi} \di z \; F(z) \equiv I_{\Gamma_\varphi} \frac{x}{z}
$$
where $\Gamma_\varphi$ is the ${\rm arg} z = \varphi$ line on the complex plane. Therefore, for
fixed $\varphi$ we have, by using the Theorem \ref{th1} and the above results:
$$
 G(z) \equiv \sum_{n=1} F(nz) = \sum_{n=1} F(nx \, \e^{\ii \varphi}) = \sum_{n=1}
  f(nx) \sim \frac{I_f}{x} + \sum_{n=0}^\infty a_n \zeta(-n) x^n = 
  \frac{I_{\Gamma_\varphi}}{z} + \sum_{n=0}^\infty A_n \zeta(-n) z^n  
$$
which concludes the proof. Now, according to the definition of distillate, if the asymptotic
expansion of $G(z)$ is convergent, we can conclude that:
$$
  \dist_0 \sum_{n=0}^\infty F(nz) = \dist_0 G(z) = \sum_{n=0}^\infty A_n \zeta(-n) z^n .
$$
The integral $I_{\Gamma_\varphi}$, in general, depends on the phase of the complex variable $z$ 
by construction. According to the definition of distillate then, under the same hypothesis of 
convergence:
$$
  \dist_0 z^M G(z) = \dist_0 \left( I_{\Gamma_\varphi} z^{M-1} + \sum_{n=0}^\infty A_n \zeta(-n) 
  z^{n+M} \right) = \sum_{n=0}^\infty A_n \zeta(-n) z^{n+M}.
$$
However, if the function $F(z)$ is analytic in the complex plane, including $z=0$, then the
integral $I_{\Gamma_\phi}$ does not depend on the integration path because the integral on the arc $\Gamma_R$ vanishes for 
$R \to \infty$ owing to the hypothesis of fast decay of $f$ in the theorem (see figure~\ref{fig:paths}). 
In this case, the asymptotic expansions is independent of the domain or angular sector and we have:
%
$$
  \dist_0 z^M G(z) = I_{\mathbb R^+} z^{M-1} + \sum_{n=0}^\infty A_n \zeta(-n) z^{n+M},
$$
where
$$
I_{\mathbb R^+} = I_{\Gamma_\varphi} =  \int_0^\infty \di z \; F(z).
$$

The Theorem \ref{th1} has an important extension to asymptotic expansions with negative powers 
\cite{zagier}:
\begin{theorem}\label{th2}
{\em Let $f: (0,+\infty) \to \mathbb{C}$ be a $C^\infty$ complex valued function on the positive real 
axis with the asymptotic power series in $x=0$ 
$$
  f(x) \sim \sum_{n=-M}^\infty a_n x^n
$$  
and $f(x)-a_{-1}/x = {\cal O}(1/x^{1+\epsilon})$ for $x \to +\infty$ with $\epsilon > 0$. 
Then, the function defined by the series:
$$
   g(x) = \sum_{n=1}^\infty f(nx)
$$
has the asymptotic expansion for $x \to 0^+$:
$$
   g(x) \sim \frac{1}{x}\left( - a_{-1} \log x + I_f \right) 
   + \sum_{\substack{n=-M \\ n \ne -1}}^\infty a_n \zeta(-n) x^n ,
$$   
where 
$$
  I_f = \int_0^\infty \di x \; \left( f(x)-\sum_{n=-M}^{-2}a_n x^n - a_{-1} \e^{-x}/x \right).
$$ }
\end{theorem}

It can be seen that, if $a_{-1} \ne 0$, the asymptotic expansion for $x \to 0^+$ has a
logarithmic term and it is thus not an asymptotic power series. Nevertheless, if $a_{-1}=0$,
this theorem provides, again, an asymptotic power series which is fit for analytic distillation 
according to the definition above. Furthermore, if $a_{-1}=0$, the Theorem \ref{th2} can be extended 
to complex variables with the same argument following Theorem \ref{th1} 
\be\label{complexg}
 G(z) \sim \frac{I_{\Gamma_\varphi}}{z} + \sum_{\substack{n=-M \\ n \ne -1}}^\infty A_n \zeta(-n) z^n  
\ee
with:
$$
  I_{\Gamma_\varphi} = \int_{\Gamma_\varphi} \di z \; \left( F(z)-\sum_{n=-M}^{-2} A_n z^n \right).
$$
Again, if the function $F(z)-\sum_{n=-M}^{-2} A_n z^n$ is analytic, the integral can be done on
any path and no dependence on the argument of $z$ arises in the asymptotic expansion in the 
complex plane.

\section{Study of the series and comparison with known results: acceleration}
\label{sec:accel}

We are now going to study a special case of global equilibrium which is obtained by setting 
the constant vector $b$ and the antisymmetric tensor $\varpi$ in eq.~\eqref{general} to
\be\label{acc}
 b_\mu = (1/T_0,0,0,0), \qquad \qquad
 \varpi_{\mu\nu} = (a/T_0) (g_{0\nu} g_{3\mu} - g_{3\nu} g_{0\mu}),
\ee
with $T_0$ and $a$ real positive constants. The resulting density operator can be written:
\be\label{accdo}
  \wrho = \frac{1}{Z} \exp \left[ -\widehat H /T_0 + a \widehat K_z/T_0 \right]
\ee
with $\widehat K_z \equiv \widehat J_{30}$ being the generator of a Lorentz boost along the 
$z$ axis. The combination $\widehat H - a \widehat K_z$ can be seen as the generator of 
translations along its flow lines \cite{leinaas}. This form, and its relation with the Unruh 
effect, has been recently studied in some detail in refs.~\cite{Becattini:2017ljh,Becattini:2019poj,Prokhorov:2019cik,Prokhorov:2019hif,Prokhorov:2019yft}; 
we only mention here that the constant $a$ is the constant acceleration of the comoving observer with 
velocity $u= \beta/\sqrt{\beta^2}$ with a hyperbolic world-line through the origin and $T_0$
the temperature measured by a comoving thermometer with the same world-line.

From the equation~\eqref{acc} we get:
\be\label{accfields}
  b = \frac{1}{T_0}(1,{\bf 0}), \qquad \qquad \varpi \cdot x = \frac{a}{T_0} (z,0,0,t),
  \qquad \quad \beta = \frac{1}{T_0}(1+az,0,0,at).
\ee
In this case, there is a point $x_0 = (0,0,0,-1/a)$ such that: 
$$
  b = - \varpi \cdot x_0 .
$$
Hence, from eqs.~\eqref{useful2} and \eqref{btilde}:
\be\label{tildebacc}
 \tilde b = - \ii ({\sf I} - \e^{a/T_0 {\sf K}_z})x_0
\ee 
and, because of the \eqref{tildebeta2}:
\be\label{tildebetaacc}
 \tilde \beta =  \ii ({\sf I} - \e^{a/T_0 {\sf K}_z})(x - x_0).
\ee
Note that:
\begin{equation*}
{\sf I} - \e^{a/T_0 {\sf K}_z}
= \left(\begin{array}{cccc}
    1 - \cos (a/T_0) & 0 & 0 & - \ii \sin(a/T_0) \\
     0 & 0 & 0 & 0          \\
     0 & 0 & 0 & 0          \\
- \ii \sin (a/T_0) & 0 & 0 & 1 - \cos (a/T_0)  
\end{array}\right)
\end{equation*}
so that if $a/T_0 = 2\pi$, at the Unruh temperature, the above matrix vanishes and so does 
$\tilde \beta$, making the distribution function $f_c(x,p)$ in \eqref{phspacec4} independent 
of $x$ and $p$, though divergent.

An important result for global equilibrium \eqref{general} is that the mean values of local
operators depend on $x$ only through the four-temperature vector, that is \cite{Becattini:2019poj}:
\be\label{xindep}
   \media{ \widehat O(x) }_{b,\varpi} = \media{ \widehat O(0)}_{\beta(x),\varpi}
\ee
where the mean value is calculated with the density operator \eqref{general} with the parameters 
in the subscript and $\beta(x)$ given by the eq.~\eqref{fourtemp}. For instance, for a 
scalar operator, this implies that it may depend only on the scalars that can be formed
with $\beta$ and its only non-vanishing derivative, that is $\varpi$. Looking at the equation
\eqref{thvort},\eqref{alphaw}, they are:
\be\label{lorscal}
  \beta^2, \qquad \alpha^2, \qquad w^2, \qquad \alpha \cdot w .
\ee
However, in the pure acceleration case, we have $w=0$ and we are just left with two independent
scalars, $\beta^2$ and $\alpha^2$; moreover, $\alpha^2 = a^2/T^2$ turns out to be uniform throughout
\cite{Becattini:2017ljh}. Thus, once a scalar function is calculated in $x=0$, it can be inferred everywhere
by replacing $\beta^2(0)$ with $\beta^2(x)$. Therefore, the calculation of 
$\media{ \widehat O(0)}_{\beta(0)=b,\varpi}$ is sufficient to determine the whole function. 
For vector or general tensor fields, this conclusion holds, because it is possible to reduce the
calculation of a vector or a tensor field to a set of scalar fields once decomposed onto the 
vectors $\beta$, the tensor $\varpi$ and the metric tensor $g$. Altogether, in the pure acceleration 
case, we can determine every local mean value by setting $x=0$ in the distribution function and study 
the series:
$$
  f_c(0,p) =   \frac{1}{(2\pi)^3} \sum_{n=1}^\infty \exp[ - n \tilde b(-n \varpi) \cdot p],
$$
where, from the \eqref{tildebacc}:
\be\label{tildebacc2}
 \tilde b = \frac{1}{a} \left(\sin (a/T_0),0,0, \ii (1-\cos (a/T_0)) \right).
\ee
The resulting series is:
\be\label{phspace-a}
  f_c(0,p) = \frac{1}{(2\pi)^3} \sum_{n=1}^\infty 
  \exp \left[ - \frac{1}{a} \sin (n a/T_0) \varepsilon 
   \right] \exp \left[  \frac{\ii}{a} (1 - \cos(n a/T_0)) p_z \right] ,
\ee
where $\varepsilon$ is the energy and $p_z$ the contravariant component of the momentum along 
the $z$ direction. The above series is not convergent for $a \ne 0$ because the limit of the 
sequence is undefined. Note that the above series, as it stands, is not an asymptotic series 
because the terms are not an asymptotic sequence for $a \to 0$ according to the
definitions~\cite{erdelyi}.

Since the simple replacement of the real acceleration in the phase space distribution function 
does not give rise to a convergent series, one may wonder whether an analytic continuation can
be obtained through different methods, like e.g. Borel resummation. However, this would not 
cure the problem of \eqref{phspace-a}. Moreover, we should consider that the non-convergence
of the series in the physical case could arise from non-analytic terms in $\varpi=0$, specifically 
$a/T_0=0$, which are hidden in the form \eqref{phspace-a}, as discussed in some detail in Section \ref{math}. 
Indeed, since in the acceleration case a non-vanishing vector $v =-\ii x_0 = (0,0,0,\ii/a)$ exists 
such that:
$$
 \tilde b = ({\sf I}-{\sf \Lambda}(\ii a/T_0)) v
$$
according to the formula \eqref{tildebacc}, there are non-trivial solutions of the homogeneous 
equation \eqref{homog} and an instance can be easily found which is actually non-analytic for $a=0$ 
(see Appendix \ref{solhom}). 
Therefore, the idea is to single out the physical solution through the method of analytic distillation 
in $a=0$. We will not apply this procedure to the distribution function \eqref{phspace-a} itself, but to some
integrals thereof, because a comparison can be done with calculations of thermal expectation values carried 
out with a completely independent method. 

\subsection{Distillation and comparison with known results}

Quantum field theory at thermodynamic equilibrium with the density operator \eqref{accdo} can be 
approached by solving field equations in Rindler coordinates and calculating the expectation values 
of relevant creation and annihilation operators. Since Wigner function is defined with a normal ordering,
all expressions obtained in Section \ref{phasespace} and, consequently, the distribution function 
\eqref{phspacec4} involve a normal ordering in Cartesian coordinates. Therefore, we should obtain 
the same thermal expectation values in Rindler-based calculations with the subtraction of the Minkowski
vacuum contribution; particularly, all obtained results should vanish when $a/T_0=2\pi$, that
is at the Unruh temperature~\cite{Becattini:2017ljh}.

To start with, we will compare the mean value of $\wpsi^\dagger \wpsi$ for a single real
uncharged scalar field in the massless neutral case  (hence $\zeta=0$) which has been
calculated analytically in Rindler coordinates \cite{Becattini:2019poj}. 
According to the \eqref{wigner},\eqref{wigner2} and the definition of $f_c(x,p)$ in \eqref{phspacec},
we have:
$$
  \langle : \wpsi^2(x) : \rangle = \frac{1}{2} \int \di^4 k \; W(x,k)
  = \int \frac{\di^3 {\rm p}}{\varepsilon} \; f_c(x,p) 
$$
and from now on we again set $\tilde\beta_n \equiv \tilde \beta(-n\varpi)$. Note that the momentum integral
of $f_c(x,p)$ is real, which can be proved from the definition \eqref{phspacec} taking advantage of
the hermiticity of the density operator. As has been mentioned, it suffices to calculate the above 
integral at $x=0$, and yet the series \eqref{phspace-a} expressing $f_c$ is divergent in the physical 
case. However, we can write the integrand as a series for imaginary acceleration, that is setting 
$a/T_0 = - \ii \phi$ and obtain the analytically continued mean value of the field squared 
$\langle : \wpsi^2(x) : \rangle_I$:
\be\label{psi2}
  \langle : \wpsi^2(x) : \rangle_I =
  \frac{1}{(2\pi)^3} \int \frac{\di^3 {\rm p}}{\varepsilon} \sum_{n=1}^\infty \exp[ - n \tilde \beta_n \cdot p]
\ee
where, by using \eqref{tildebacc2}:
\begin{equation}\label{eq:betanAcc}
\tilde{\beta}_n(0)= \tilde b_n = \left(\frac{\sinh (n\phi)}{n T_0\phi},0,0,
    -2\frac{\sinh^2(n\phi/2)}{n T_0\phi} \right),\quad
\tilde{\beta}_n(0)^2= \tilde\beta_n \cdot \tilde{\beta}_n = \frac{4 \sinh^2(n\phi/2)}{n^2 T_0^2 \phi^2},
\end{equation}
where the scalar product is the usual Minkowskian between complex vectors. Note that $\tilde b_n$ is indeed 
real and future time-like for any real $\phi$. With such a vector, 
the series in \eqref{psi2} is uniformly convergent for real values of $\phi$ and can be integrated term by 
term. The result is, for a massless field:
\begin{equation}\label{eq:MasslesMomInt}
\int\frac{\di^3 \p}{\varepsilon}\, \e^{- n \tilde \beta_n \cdot p}
 =\frac{4 \pi}{n^2 (\tilde\beta_n \cdot \tilde \beta_n)};
\end{equation}
therefore:
$$
  \langle : \wpsi^2(x) : \rangle_I =  \frac{1}{2\pi^2} \sum_{n=1}^\infty
  \frac{1}{n^2 \tilde\beta_n \cdot \tilde{\beta}_n} = \frac{T_0^2 }{8\pi^2} \sum_{n=1}^\infty  
  \frac{\phi^2}{\sinh^2(n\phi/2)}
$$
which has the correct known limit of $T_0^2/12$ for $\phi = 0$.

We now have to go back to the physical case, i.e. to continue the above expressions to imaginary $\phi$ or 
real accelerations of the function defined by the series:
\begin{equation}\label{eq:S2series}
  S_2(\phi) = \sum_{n=1}^\infty  \frac{\phi^2}{\sinh^2(n\phi/2)} = \phi^2 G_2(\phi).
\end{equation}
However, as we have pointed out right before this subsection, the above function may include 
non-analytic contributions which should be subtracted away through the procedure of analytic
distillation in $\phi=0$ described in Section \ref{math}. Indeed, the function $S_2(\phi)$ is peculiar on 
the complex plane; it is analytic if ${\rm Re} \, \phi \ne 0$ because it can be readily shown that 
the series is uniformly convergent, whereas for imaginary $\phi$ one has:
$$
   \sum_{n=1}^\infty  \frac{({\rm Im}\phi)^2}{\sin^2(n{\rm Im}\phi/2)},
$$
which is certainly divergent as $1/\sin^2(n\, {\rm Im}\phi/2) \ge 1$; thus, the function $S_2(\phi)$
is non-analytic everywhere on the imaginary axis, although it is convergent for $\phi=0$. Moreover,
the series is not an asymptotic series for $\phi \to 0$ \cite{erdelyi} and cannot even be resummed
with the Borel method because zeroes of the denominators are overall dense.
Nevertheless, for real $\phi$, the series $G_2(\phi)$ fulfills the requirements of the Theorem \ref{th2} of 
Section~\ref{math} with $a_{-1}= 0$, hence an asymptotic power series around $\phi =0$ can be found 
which can be extended in the complex plane according to the discussion following Theorem \ref{th2} and
the equation \eqref{complexg}. Applying the Theorem~\ref{th2} one finds, being $G_2(\phi)$ an even function:
$$
G_2(\phi)\sim \begin{cases} \frac{2\pi^2}{3\phi^2} + \frac{1}{6} - \frac{2}{\phi} &  {\rm Re}\, \phi >0\\
    \frac{2\pi^2}{3\phi^2} + \frac{1}{6} + \frac{2}{\phi} & {\rm Re}\, \phi <0 \end{cases}.
$$
Note that the asymptotic power series of $G_2(\phi)$ is finite, because of the vanishing of the
$\zeta(-n)$ for even $n$ in the \eqref{complexg}. This makes it possible to make an analytic distillation 
in $\phi=0$ of the function $S_2$:
$$
\dist_0 S_2(\phi) = \dist_0 \phi^2 G_2(\phi) = \frac{2\pi^2}{3} + \frac{\phi^2}{6}.
$$
The physical solution is then obtained by setting $\phi = \ii a/T_0$:
\begin{equation*}
  \langle : \wpsi^2 (0): \rangle = \frac{T_0^2}{8 \pi^2} \left( \frac{2 \pi^2}{3} - \frac{a^2}{6 T_0^2} \right)
\end{equation*}
and, taking into account that $\beta^2(0) = 1/T_0^2$ and $a^2/T_0^2 = -\alpha^2$, we have, in a
generic point $x$:
\be\label{psi2-2}
  \langle : \wpsi^2 (x): \rangle =  \frac{1}{\beta(x)^2}
   \left(\frac{1}{12} + \frac{\alpha^2(x)}{48\pi^2}\right).
\ee
This result coincides with the exact solution in Rindler coordinates found in 
refs.~\cite{Becattini:2017ljh,Becattini:2019poj} with subtraction of Minkowski vacuum contribution.
Indeed, as we have pointed out in Section~\ref{phasespace}, the normally ordered mean value of
an operator which is a quadratic combination of creation and annihilation operators of momentum
eigenstates coincides with the mean value with subtraction of the vacuum expectation value. 
Therefore we expect, according to the general features of the Unruh effect and the conclusions
of ref.~\cite{Becattini:2017ljh}, the normally ordered thermal expectation value \eqref{psi2-2}, 
as well as any other quadratic operator in the fields to vanish precisely at the Unruh temperature, 
that is $a/T_0 = \sqrt{-\alpha^2} = 2\pi$ and this actually occurs. Furthermore, since \eqref{psi2} 
is an integral of the complex distribution function $f_c(x,p)$, we argue that this characteristic 
must extend to {\em any} integral of the distribution function after analytic distillation and 
continuation; this will be demonstrated for a large class of integrals in the Appendix \ref{distillac}.
As pointed out below eq.~\eqref{tildebetaacc}, for $a/T_0 = 2\pi$ the equation \eqref{tildebetaacc} yields
$\tilde \beta = 0$, hence, for $\zeta=0$ a complex distribution function \eqref{phspacec4} which is 
a divergent constant independent of $x$ and $p$. 
\vspace{0.4cm}
\begin{center}
    \rule{10cm}{0.4pt}
\end{center}
\vspace{0.4cm}
In order to confirm the agreement found, we have compared the expectation values of the 
canonical stress-energy tensor with an exact calculation in Rindler coordinates, reported in 
Appendix \ref{exactrind}. If the distribution function were known, the stress-energy tensor
could be calculated using the eq.~\eqref{cantens2}, which select only the real part of
a momentum integral. But we only know the distribution function through eq.~\eqref{phspacec4}
that is only valid for imaginary value of thermal vorticity. Therefore to obtain the physical
expectation values we have to compute the momentum integrals by plugging the distribution
function representation~\eqref{phspacec4} into~\eqref{cantens2} for imaginary thermal vorticity,
then continue it to physical values and at last take the real part:
\be\label{eq:setRealP}
\langle : \wT_C^{\mu\nu}: \rangle
 = {\rm Re} \left[\langle : \wT_C^{\mu\nu}: \rangle_I\right]_{\phi=\ii a/T_0},
\ee
where the imaginary expectation value is the series:
\be\label{stress1}
 \langle : \wT_C^{\mu\nu}: \rangle_I = \frac{1}{(2\pi)^3} \sum_{n=1}^\infty 
  \left[ \int \frac{\di^3 p}{\varepsilon} \; \left( p^\mu p^\nu + \frac{1}{4} 
  \left( \ii p^\mu \partial^\nu + \ii p^\nu \partial^\mu \right) +
  \frac{1}{4} \left( \partial^\mu \partial^\nu - g^{\mu\nu} \Box \right) \right) 
  \e^{-n \tilde\beta_n \cdot p}\right].
\ee
The momenta in the integrand of eq.~\eqref{stress1} can be expressed as derivatives with respect 
to $\tilde\beta_n$:
\begin{equation*}
\begin{split}
 \left( p^\mu p^\nu + \frac{1}{4} \left( \ii p^\mu \partial^\nu + \ii p^\nu \partial^\mu \right) +
  \frac{1}{4} \left( \partial^\mu \partial^\nu - g^{\mu\nu} \Box \right) \right) 
  \e^{-n \tilde\beta_n \cdot p} = & \left( \frac{1}{n^2}
   \frac{\partial}{\partial \tilde \beta_\mu}\frac{\partial}{\partial \tilde \beta_\nu} 
  - \frac{1}{4n} \left( \ii \partial^\nu \frac{\partial}{\partial \tilde \beta_\mu} + 
  \ii \partial^\mu \frac{\partial}{\partial \tilde \beta_\nu} \right)+ \right.\\
  &\left.+
  \frac{1}{4} \left( \partial^\mu \partial^\nu - g^{\mu\nu} \Box \right) \right) 
  \e^{-n \tilde\beta_n \cdot p}.
\end{split}
\end{equation*}
Now the momentum integral is the same as eq.~\eqref{eq:MasslesMomInt} and we obtain:
\be\label{stress2}
 \begin{split}
 \langle : \wT_C^{\mu\nu}: \rangle_I = & \frac{1}{(2\pi)^3} \sum_{n=1}^\infty 
  \left( \frac{1}{n^2}
   \frac{\partial}{\partial \tilde {\beta_n}_\mu}\frac{\partial}{\partial \tilde {\beta_n}_\nu} 
  - \frac{1}{4n} \left( \ii \partial^\nu \frac{\partial}{\partial \tilde {\beta_n}_\mu} + 
  \ii \partial^\mu \frac{\partial}{\partial \tilde {\beta_n}_\nu} \right) +
  \frac{1}{4} \left( \partial^\mu \partial^\nu - g^{\mu\nu} \Box \right) \right)\times\\
  &\times\frac{4\pi}{n^2 (\tilde\beta_n\cdot \tilde\beta_n)}.
  \end{split}
\ee
The derivatives can be worked out by using:
\begin{equation*}
\begin{split}
\frac{\partial}{\partial \tilde \beta_\mu}\frac{1}{\tilde\beta \cdot \tilde \beta} &=
	-2\frac{\tilde\beta^\mu}{\tilde\beta \cdot \tilde \beta}\frac{1}{\tilde\beta \cdot \tilde \beta};\qquad
\frac{\partial}{\partial \tilde \beta_\mu}\frac{\partial}{\partial \tilde \beta_\nu} \frac{1}{\tilde\beta \cdot \tilde \beta} =
	\left( 8\frac{\tilde\beta^\mu \tilde\beta^\nu}{(\tilde\beta \cdot \tilde \beta)^2}
    	-2\frac{g^{\mu\nu}}{\tilde\beta \cdot \tilde \beta}\right)\frac{1}{\tilde\beta \cdot \tilde \beta};\\
\partial^\nu \frac{\partial}{\partial \tilde \beta_\mu}\frac{1}{\tilde\beta \cdot \tilde \beta} &=
	\left( 8\frac{\tilde\beta^\mu \tilde\beta_\rho \partial^\nu\tilde\beta^\rho}{(\tilde\beta \cdot \tilde \beta)^2}
    	-2\frac{\partial^\nu\tilde\beta^\mu}{\tilde\beta \cdot \tilde \beta}\right)\frac{1}{\tilde\beta \cdot \tilde \beta};\\
\end{split}
\end{equation*}
and one eventually obtains:
\be\label{setfinal}
\media{:\wT_C^{\mu\nu}(x):}_I=\frac{1}{2\pi^2}\sum_{n=1}^\infty
  \left[ t^{\mu\nu}(n)+t^{\nu\mu}(n)-g^{\mu\nu}t(n) \right],
\ee
where:
\be\label{deftmunueet}
\begin{split}
t^{\mu\nu}(n)&=\left(\frac{4}{n^2}\frac{\tilde\beta_n^\mu \tilde\beta_n^\nu}{(\tilde\beta_n \cdot \tilde \beta_n)^2}
	-\frac{2\ii}{n}\frac{\tilde\beta_n^\mu \tilde\beta_{n\rho} \partial^\nu\tilde\beta_n^\rho}{(\tilde\beta_n \cdot \tilde \beta_n)^2}
    +\frac{\ii}{2n}\frac{\partial^\nu\tilde\beta_n^\mu}{\tilde\beta_n \cdot \tilde \beta_n}
    +\frac{1}{8}\partial^\mu \partial^\nu\right)\frac{1}{n^2(\tilde\beta_n \cdot \tilde \beta_n)}\\
t(n)&=\left(\frac{2}{n^2}\frac{1}{\tilde\beta_n \cdot \tilde \beta_n}+\frac{1}{4}\Box\right)\frac{1}{n^2(\tilde\beta_n \cdot \tilde \beta_n)}.
\end{split}
\ee
As has been pointed out earlier in this section (see eq.~\eqref{xindep}), it is sufficient to 
calculate the stress-energy tensor in $x=0$. From the \eqref{tildebetaacc} one can write the vector 
field $\tilde\beta_n(x)$ as a function of the Minkowskian coordinates $t$ and $z$:
\be\label{tildebetaaccn}
\tilde{\beta}^\mu_n(x)=\frac{1}{n}\left(
\ii t (\cosh(n\phi)-1)+\frac{\sinh(n\phi)\left(\frac{1}{T_0}+\ii z\phi\right)}{\phi},0,0,
\ii t \sinh (n \phi )+\frac{(\cosh (n \phi )-1) (1+\ii T_0 z \phi )}{T_0 \phi } \right),
\ee
whence:
\begin{equation*}
\tilde\beta_n(x) \cdot \tilde \beta_n(x) =-\frac{4 \sinh^2\left(\frac{n \phi}{2}\right)
    \left(T_0^2 \phi^2 \left(z^2-t^2\right)-2 \ii T_0 z \phi -1\right)}{n^2 T_0^2 \phi^2},
\end{equation*}
and:
\begin{equation*}
\de^\mu \tilde{\beta}^\nu_n(x) = \left(
\begin{array}{cccc}
 \frac{\ii (\cosh (n \phi )-1)}{n} & 0 & 0 & \frac{i \sinh (n \phi )}{n} \\
 0 & 0 & 0 & 0 \\
 0 & 0 & 0 & 0 \\
 -\frac{\ii \sinh (n \phi )}{n} & 0 & 0 & -\frac{i (\cosh (n \phi )-1)}{n} \\
\end{array}
\right) .
\end{equation*}
The derivatives of the $\tilde \beta_n$ needed to work out the \eqref{deftmunueet} in $x=0$ 
are:
\begin{equation*}
\left(\de^\mu \de^\nu\frac{1}{\tilde\beta_n \cdot \tilde \beta_n}\right)\Bigg|_{x=0}
=\text{diag}\left(
\frac{n^2 T_0^4 \phi ^4}{1-\cosh (n \phi )},
0,0,\frac{3 n^2 T_0^4 \phi ^4}{1-\cosh (n \phi )}\right).
\end{equation*}
From the equations \eqref{setfinal},\eqref{deftmunueet} and using the above derivatives
we get:
\begin{equation}\label{setzero}
\media{:T^{\mu\nu}(0):}_I=\frac{1}{16 \pi^2} \sum_{n=1}^\infty\left(\begin{array}{cccc}
3T_0^4\phi^4 \sinh^{-4}\left( n\phi/2 \right) & 0 & 0 & 0 \\
0 & T_0^4\phi^4 \frac{\cosh(n\phi)}{\sinh^4\left(n\phi/2 \right)} & 0 & 0 \\
0 & 0 & T_0^4\phi^4 \frac{\cosh(n\phi)}{\sinh^4\left(n\phi/2 \right)} & 0 \\
0 & 0 & 0 & T_0^4\phi^4\sinh^{-4}\left(n\phi/2 \right) \\
\end{array}\right).
\end{equation}

Because of the symmetries of the density operator \eqref{accdo} and the fact that the 
stress-energy tensor can only depend on $\beta$ and $\varpi$ and not explicitely on $x$, 
according to the discussion about equation \eqref{xindep}, it may only have this form 
\cite{Becattini:2019poj}:
\begin{equation}\label{eq:setAcc}
\media{:T^{\mu\nu}(x):} = \rho\, u^\mu u^\nu - p\,(g^{\mu\nu}-u^\mu u^\nu) 
	 + {\cal A} \, \alpha^\mu \alpha^\nu,
\end{equation}
where $u^\mu = \beta^\mu/\sqrt{\beta^2}$ and the thermal functions $\rho,\,p$ and $\cal A$ can only 
depend on the Lorentz scalars $\beta^2$ and $\alpha^2$.
From the \eqref{setzero} and the~\eqref{eq:setRealP} we then get the physical thermal functions:
\begin{equation*}
\begin{split}
\rho & =\frac{1}{16 \pi^2} \text{Re}\left[\sum_{n=1}^\infty 3T_0^4\phi^4 \sinh^{-4}\left( n\phi/2 \right)\right]_{\phi=\ii a/T_0},\\
p &=\frac{1}{16 \pi^2} \text{Re}\left[\sum_{n=1}^\infty T_0^4\phi^4 \frac{\cosh(n\phi)}{\sinh^4\left(n\phi/2 \right)}\right]_{\phi=\ii a/T_0},\\
{\cal A} &=\frac{1}{16 \pi^2}\text{Re}\left[\sum_{n=1}^\infty 2T_0^4\phi^4 \sinh^{-2}\left( n\phi/2 \right)\right]_{\phi=\ii a/T_0}.
\end{split}
\end{equation*}
The series are similar to the one in eq.~\eqref{eq:S2series} and they can be tackled in
the same manner. The analytic distillation and the continuation to physical acceleration with
$a = - \ii T_0 \phi$ of these series are discussed in Appendix~\ref{distillac}. The result is:
\begin{equation*}
\begin{split}
\rho & = T_0^4 \left( \frac{\pi^2}{30} + \frac{a^2}{12 T_0^2} -\frac{11 a^4}{480\pi^2 T_0^4}\right),\\
p &= T_0^4 \left(\frac{\pi^2}{90} - \frac{a^2}{18 T_0^2} + \frac{19 a^4}{1440\pi^2 T_0^4} \right),\\
{\cal A} &= T_0^4 \left(\frac{1}{12} - \frac{a^2}{48 \pi^2 T_0^2} \right).
\end{split}
\end{equation*}
The form of the thermodynamic coefficient in any point $x$ can be inferred from the previous 
result simply by taking into account that $\beta^2(0) = 1/T_0^2$ and $a^2/T_0^2 = -\alpha^2$, 
so we have:
\begin{equation}\label{eq:ResAcc}
\begin{split}
\rho & =\beta(x)^{-4}\left(\frac{\pi^2}{30}-\frac{\alpha^2(x)}{12}-\frac{11\alpha^4(x)}{480\pi^2}\right),\\
p &=\beta(x)^{-4}\left( \frac{\pi^2}{90}+\frac{\alpha^2(x)}{18}+\frac{19\alpha^4(x)}{1440\pi^2}\right),\\
{\cal A} &=\beta(x)^{-4}\left(\frac{1}{12}+\frac{\alpha^2(x)}{48 \pi^2}\right).
\end{split}
\end{equation}
These coefficients nicely coincide with those calculated by solving field equations in Rindler
coordinates in the right Rindler wedge (reported in Appendix~\ref{exactrind}), where the four-temperature 
is a time-like vector. Moreover, they all vanish at the Unruh temperature $T_0=a/2\pi$ that is for
$|\alpha|=2\pi$, as expected for a quadratic operator according to the discussion following 
equation \eqref{psi2-2}. 

These results precisely coincide with a perturbative expansion in acceleration to quadratic order 
in refs.~\cite{Buzzegoli:2017cqy} and to fourth order presented in refs.~\cite{Prokhorov:2019cik,Prokhorov:2019hif} 
for the same statistical operator in \eqref{accdo}. This result bears out the observation made
in refs.~\cite{Prokhorov:2019cik} that the vanishing of the stress-energy tensor at the Unruh 
temperature is achieved, in the perturbative expansion in $a/T$ at the fourth-order which led 
the same authors to conclude that the exact expression must be polynomial \cite{Prokhorov:2019hif},
namely that the \eqref{eq:ResAcc} were indeed the complete solutions.
These authors also pointed out that these same results can be derived within a non-perturbative
approach in a quantum field theory over a space with a conical singularity \cite{Prokhorov:2019yft}, 
see for instance~\cite{Dowker:1977zj} and other references in~\cite{Prokhorov:2019yft}. 

\section{Study of the series and comparison with known results: rotation}
\label{rotation}

We are now going to study a special case of global equilibrium which corresponds to system rigidly rotating
along the $z$ axis and which is obtained from eq.~\eqref{general} by setting:
\be\label{rot}
 b_\mu = (1/T_0,0,0,0), \qquad \qquad
 \varpi_{\mu\nu} = (\omega/T_0) (g_{1\mu} g_{2\nu}  - g_{1\nu} g_{2\mu}),
\ee
where $\omega$ and $T_0$ are positive constants. More in general, without specifing the axis of rotation, we
can choose
$$
  b = \frac{1}{T_0} (1,{\bf 0}), \qquad \qquad \varpi \cdot x = \frac{1}{T_0} (0,\omegav \times {\bf x} ),
$$
with $\omegav$ constant vector. It can be shown that $\omega$ has the physical meaning of 
a constant angular velocity \cite{landau} and $T_0$ is the temperature measured on the axis
of rotation. This state is also known as {\em rotational thermodynamic equilibrium}. The 
density operator becomes:
\be\label{rotdo}
  \wrho = \frac{1}{Z} \exp \left[ -\frac{\widehat H}{T_0} + \frac{1}{T_0} \omegav \cdot \widehat{\bf J} \right].
\ee
Because of equation \eqref{btilde}, it turns out that $\tilde b (\varpi) = b$. In the physical 
case with real $\varpi$, that is real $\omegav$, the eq.~\eqref{tildebeta2} can be rewritten as:
$$
  \tilde \beta (\varpi) =  b - \ii\left( {\sf I} - {\sf R}(\ii \omegav/T_0) \right)x,
$$
where ${\sf R}(\ii \omegav/T_0) = \exp[\omegav \cdot {\bf J}/T_0]$ is the rotation around 
the axis defined by $\omegav$ by an imaginary angle $\ii\omega/T_0$. Thus:
\begin{align}\label{tildebetarot}
 \tilde \beta(\varpi) & = b + \ii \left[ \sin (\ii \omega/T_0) {\hat\omegav \times {\bf x}} +
  (1 - \cos(\ii \omega/T_0)) \hat \omegav \times (\hat \omegav \times {\bf x}) \right]
  \nonumber \\
  &= b - \sinh (\omega/T_0) {\hat\omegav \times {\bf x}} + \ii
  (1 - \cosh(\omega/T_0)) \hat \omegav \times (\hat \omegav \times {\bf x})
\end{align}
and, consequently:
$$
  \tilde \beta(\varpi) \cdot p = \frac{\varepsilon}{T_0} + \sinh \left( \frac{\omega}{T_0} \right)  
  \hat\omegav \cdot ({\bf x} \times {\bf p}) - \ii \left( 1 - \cosh \left( \frac{\omega}{T_0} 
  \right) \right) (\hat \omegav \cdot {\bf x}\, \hat \omegav - {\bf x}) \cdot {\bf p}.
$$
Therefore, the analytically continued distribution function is, from eq.~\eqref{phspacec4}:
\be\label{phsprot}
 f(x,p) = {\rm Re} \frac{1}{(2\pi)^3} \sum_{n=1}^\infty \exp \left[ - n \frac{\varepsilon}{T_0} + 
 n \sinh \left( \frac{n \omega}{T_0} \right)  \hat\omegav \cdot ({\bf x} \times {\bf p}) +  
  n \ii \left( 1 - \cosh \left( \frac{n \omega}{T_0} \right) \right) (\hat \omegav \cdot {\bf x} 
  \,\hat \omegav - {\bf x}) \cdot {\bf p} + \zeta n \right]
\ee
which converges only if $({\bf x} \times {\bf p})\cdot \hat \omegav < 0$, that is only if
the orbital angular momentum along the $\omegav$ direction is negative. Otherwise, the series
is divergent, which seems to imply that there is an infinite number of particles with 
$({\bf x} \times {\bf p})\cdot \hat \omegav > 0$.

To understand the source of this problem, and to prove that the distribution function found
is the actual exact solution, it is very convenient to diagonalize the density operator 
\eqref{rotdo}, following the method of ref.~\cite{vilenkin}. Setting $\hat \omegav = \hat{\bf k}$ 
as the direction of the $z$ axis, this can be done by using single particle states with 
eigenvalues $p_T,p_z$  (transverse and longitudinal momentum) and the eigenvalue of the angular 
momentum $M$ which takes on integer values. The relation between momentum eigenstates and 
eigenstates $\ket{p_T,p_z,M}$ is well known and leads to the relations between creation operators:
\be\label{arelaz}
   \wad{p} = \sqrt{2 \varepsilon} 
   \sum_{M=-\infty}^\infty \frac{1}{\sqrt{2\pi}} \e^{-\ii M \varphi} \wad{p_T,p_z,M}
\ee
along with its inverse:
\be\label{arelazinv}
   \wad{p_T,p_z,M} = \frac{1}{\sqrt{2\pi} \sqrt{2 \varepsilon}}\int_0^{2\pi} \di \varphi \;
    \e^{\ii M \varphi} \wad{p},
\ee
where $\varphi$ is the azimuthal coordinate of ${\bf p}$. One can then plug the \eqref{arelaz} 
in the field expansion in cartesian coordinates~\eqref{field} and obtain the corresponding 
solution in cylindrical coordinates. The operator on the exponent of \eqref{rotdo} can then
be written as:
\be\label{rotexp}
  \widehat H - \omega \wJ_z = \sum_{M=-\infty}^\infty \int \di p_T p_T \di p_z \; 
   (\varepsilon - M \omega)  \wad{p_T,p_z,M} \, \wa{p_T,p_z,M} + {H_0},
\ee
where $H_0$ is a divergent constant. Unlike the Hamiltonian, the spectrum of the operator 
\eqref{rotexp} is not bounded from below because the single particle energy can be as low
as the mass $m$, whereas the angular momentum eigenvalue $M$ can take on any value, so 
$\varepsilon- M \omegav$ has no lower bound. This causes most statistical mean to be hopeless 
divergent because the
probability of large occupation numbers is exponentially large when $\varepsilon-M \omega < 0$
which is clearly the case when $M$ is positive and the energy is low enough; this is the 
origin of the divergence of the distribution function $f(x,p)$. This problem is due to the
absence of boundary conditions for the field; if the scalar field has Dirichlet boundary
condition $\wpsi=0$ at a radius $R < 1/\omega$ the spectrum is bounded from below because 
the transverse momentum eigenvalues are discrete and depending on $M$, and the following 
inequality holds \cite{vilenkin}:
$$
 \varepsilon = \sqrt{p_z^2+ m^2 + p_T(M)^2} > M \omega.
$$

Curing the divergence inherent in the rotational equilibrium without field boundary 
conditions goes beyond the scope of this work. However, we can check that the solution 
found \eqref{final2} is the same as the known one for the basis $(p_T,p_z,M)$. 
By using the same method as for $\media{\wad{p} \, \wa{p^\prime}}$ one can calculate
$\media{\wad{p_T,p_z,M} \, \wa{p_T^\prime,p_z^\prime,M^\prime}}$ with the density
operator \eqref{rotdo}; it is a much easier task for $\widehat J_z$ is diagonal on
the eigenstates created by $\wad{p_T,p_z,M}$ \cite{vilenkin}. The result is diagonal
in the mean values:
\be\label{tevm}
   \media{\wad{p_T,p_z,M} \, \wa{p_T^\prime,p_z^\prime,M^\prime}} = 
   \frac{1}{p_T} \delta(p_T-p_T^\prime)\delta(p_z-p_z^\prime) \delta_{M M^\prime}
   \sum_{n=1}^\infty \e^{-n(\varepsilon/T_0 - M \omega/T_0)} .
\ee
Note that the series is convergent only if $\varepsilon > M \omega$, see above discussion. 
For an imaginary $\omega/T_0 = -\ii \phi$, that is for an actual Lorentz transformation, the 
sum \eqref{tevm} is always convergent and the result is:
\be\label{tevm2}
   \media{\wad{p_T,p_z,M} \, \wa{p_T^\prime,p_z^\prime,M^\prime}} = \frac{1}
   {e^{\varepsilon/T_0 + \ii M \phi}-1} \frac{1}{p_T} \delta(p_T-p_T^\prime)
   \delta(p_z-p_z^\prime) \delta_{MM^\prime}\, .
\ee
The relation \eqref{tevm} can be used to calculate $\media{\wad{p} \, \wa{p^\prime}}$
by means of the \eqref{arelaz}, for $\omega/T_0 = - \ii \phi$:
\begin{align}\label{media1}
\media{\wad{p} \, \wa{p^\prime}} &= \sum_{M,M^\prime} \frac{1}{2\pi} 
 \e^{-\ii M \varphi + \ii M^\prime \varphi^\prime} 
  \media{\wad{p_T,p_z,M} \, \wa{p_T^\prime,p_z^\prime,M^\prime}} \nonumber \\
 &= \sum_{M} \frac{1}{2\pi} \e^{-\ii M (\varphi -\varphi^\prime)} 
 \sum_{n=1}^\infty e^{-n(\varepsilon/T_0 - \ii M \phi)} \frac{1}{p_T} \delta(p_T-p_T^\prime)
 \delta(p_z-p_z^\prime) \nonumber \\
  &= \sum_{n=1}^\infty e^{-n \varepsilon/T_0} \sum_{M} \frac{1}{2\pi} \e^{-\ii M 
  (\varphi -\varphi^\prime + n \phi)} \frac{1}{p_T} \delta(p_T-p_T^\prime)\delta(p_z-p_z^\prime)
  \nonumber \\
 &= \sum_{n=1}^\infty e^{-n \varepsilon/T_0} \delta (\varphi - \varphi^\prime + n\phi) 
  \frac{1}{p_T} \delta(p_T-p_T^\prime)\delta(p_z-p_z^\prime) .
\end{align}
Let us now use take \eqref{final3} with ${\sf \Lambda} = \exp[-\ii \phi {\sf J}_z]$.
Since the rotation around the $z$ axis leaves $p_T$ and $p_z$ unchanged, the vector
${\sf \Lambda}^n {\bf p}$ has cylindrical coordinates $p_T$, $\varphi + n\phi$ and $p_z$.
Hence:
$$
 \delta^3({\sf\Lambda}^n{\bf p} - {\bf p}^\prime) =  \delta (\varphi + n \phi - \varphi^\prime) 
 \frac{1}{p_T} \delta(p_T-p_T^\prime)\delta(p_z-p_z^\prime) .
$$
Therefore, being $\tilde b = b = (1/T_0,{\bf 0})$ for rotations, as we have seen, 
the eq.~\eqref{final2} becomes:
$$
 \media{\wad{p} \, \wa{p^\prime}} = \sum_{n=1}^\infty \e^{-n\varepsilon/T_0}  
 \delta (\varphi - \varphi^\prime + n\phi) \frac{1}{p_T} \delta(p_T-p_T^\prime) 
 \delta(p_z-p_z^\prime) ,
$$
which is the same as \eqref{media1}. The proof is concluded.

A remarkable feature of the calculation with imaginary angular velocity is that the series
involved in \eqref{tevm2} is always convergent. Thereby, the problem of the divergence due to 
the lack of lower bound of the operator \eqref{rotexp} is evaded by going to imaginary $\omega$
and it is possible that a suitable analytic continuation to real $\omega$ yields finite results. 
It should be stressed, however, that the physical meaning of the expressions found is limited 
because the divergence should be cured in the physical case with different methods.

This proof has another noteworthy consequence, that it is that there is no non-trivial solution
of the homogeneous equation \eqref{homog} for the density operator \eqref{rotdo}; in other words
the \eqref{final3} is the only solution of the equation \eqref{basic}. The reason is simply 
that $\media{\wad{p} \, \wa{p^\prime}}$ is the only possible expression corresponding to
the solution \eqref{tevm2}, which is in turn the only possible expression of 
$\media{\wad{p_T,p_z,M} \, \wa{p_T^\prime,p_z^\prime,M^\prime}}$ which can be obtained 
algebraically \cite{vilenkin} in the basis diagonalizing the density operator. This is consistent 
with what we described as a sufficient condition to obtain a non-trivial solution of the 
homogeneous equation \eqref{homog} as, in the case of rotation, there is no vector $v$ such 
that $\tilde b = b = (I - {\sf \Lambda})(v)$ with $b$ as in eq.~\eqref{rot} (see Appendix \ref{solhom}).

\subsection{Comparison with known results}

We now move to the calculation of thermal expectation values similar to those of the previous 
section. First we define a suitable tetrad for the rotational equilibrium with $\hat \omega = \hat {\bf k}$. 
Keeping in mind the decomposition~\eqref{thvort}, we define the following four-vectors 
(see also~\cite{Becattini:2015nva,Buzzegoli:2018wpy}):
\begin{equation*}
u^\mu =\frac{\beta^\mu}{\sqrt{\beta^2}},\quad
\alpha^\mu=\varpi^{\mu\nu}u_\nu,\quad
w^\mu=-\frac{1}{2}\epsilon^{\mu\nu\rho\sigma}\varpi_{\nu\rho}u_\sigma,\quad
l^\mu=\epsilon^{\mu\nu\rho\sigma}w_\nu\alpha_\rho u_\sigma.
\end{equation*}
The tetrad $\{u,\alpha,w,l \}$ is an orthogonal, non normalized basis of Minkowski spacetime built 
upon the four-temperature $\beta$ and thermal vorticity $\varpi$ and can be used to decompose all 
vectors and tensor fields. Moreover, denoting with $r^2=x^2+y^2$ the distance from the rotation axis 
we have that the time component of the four-velocity $u$ reads:
\begin{equation*}
 u^0 \equiv \gamma = \frac{1}{\sqrt{1-r^2\omega^2}},
\end{equation*}
and:
\begin{equation}
\label{eq:RotScal}
\beta^2(x)=\frac{1}{T_0^2\gamma^2},\quad \alpha^2(x)=-(\gamma^2-1)\frac{\omega^2}{T_0^2},
\quad w^2(x)=-\gamma^2\frac{\omega^2}{T_0^2}.
\end{equation}
Formulae \eqref{eq:RotScal} allow to transform the widely used $r^2,\omega,T_0$ variables 
into the Lorentz invariants discussed in the previous section.
Note that, unlike in the acceleration case, in the rotational case the calculation of the thermal
expectation value of a scalar operator at $x=0$, that is $r=0$, is not sufficient to know its value
everywhere because at $r=0$ we have $\alpha^2=0$ and the dependence on one of the three independent
Lorentz scalars is lost.

We now turn to the mean value of the field squared in the massless neutral case, see eq.~\eqref{psi2}. 
With $\phi=\ii \omega/T_0$, we have
\begin{equation}
\label{eq:betanRot}
\tilde{\beta}_n \cdot \tilde{\beta}_n = \frac{1}{T_0^2} + [\sin^2(n\phi)+
  (1-\cos(n\phi))^2] \frac{r^2}{n^2} = \frac{1}{T_0^2} + 4 \sin^2 \left(\frac{n \phi}{2} 
   \right) \frac{r^2}{n^2}
\end{equation}
and the series~\eqref{psi2} becomes:
\be\label{phspint2}
  \langle : \wpsi^2 : \rangle_I =  \frac{1}{2\pi^2} \sum_{n=1}^\infty
   \frac{T_0^2}{n^2 + 4 \sin^2 \left(\frac{n \phi}{2}\right) T_0^2 r^2}.
\ee
For real $\phi$ the series is absolutely and uniformly convergent, because it is bounded by the series 
of $1/n^2$; this is an expected result according to the discussion on the convergence of the series
\eqref{tevm2}. However, for complex $\phi$ the series is densely divergent. To show this, let ${\rm Re} \, \phi$ 
be a rational multiple of $2\pi$, so that ${\rm Re}\,\phi = (N/L) 2\pi$ with $N$ and $L$ irreducible integers.
The $n$-th denominator of the series in \eqref{phspint2} vanishes when $n =K L$ with $K$ integer and if: 
\be\label{curves}
{\rm Im} \, \phi = \pm\frac{2}{n} {\rm asinh} \left( \frac{n}{2r T_0} \right)
    =\pm\frac{{\rm Re} \, \phi}{K N\pi} {\rm asinh} \left( \frac{K N}{{\rm Re} \, \phi}
    \frac{\pi}{r T_0}\right),
\ee
making the series in \eqref{phspint2} a divergent one. Since $K$ and $N$ are arbitrary integer numbers, 
the curves \eqref{curves} are dense in the complex plane. Altogether, the function 
\eqref{phspint2} cannot then be defined except on the real axis. 

The divergence of the series \eqref{phspint2} for non-real $\phi$ is evidently 
a consequence of  the divergence of the distribution function \eqref{phsprot} for $({\bf x} \times {\bf p}) 
\cdot \omegav > 0$ and its physical reason has already been discussed. Even though there is no proper 
domain of existence of the complex function \eqref{phspint2}, thus no asymptotic expansion for complex
$\phi$, we can, notwithstanding, extract a finite analytic distillate of the function defined by the 
series \eqref{phspint2} for real $\phi$ by using the theorems of the Section \ref{math}, confining
ourselves to the positive and negative real axis and keeping in mind that the physical meaning of 
the solution will be limited (this point will be resumed later on). 

We start rewriting the right hand side of \eqref{phspint2} in a form which is suitable for the 
application of the Theorem \ref{th2}. By using the auxiliary real parameter $t$ we have:
\begin{equation}\label{phspint3}
    \langle : \wpsi^2 : \rangle_I =  \frac{T_0^2}{2\pi^2} \sum_{n=1}^\infty \lim_{t \to \phi}
   \frac{\phi^2}{\phi^2 n^2 + 4 \sin^2 \left(\frac{n \phi}{2}\right)t^2 T_0^2 r^2}=\frac{T_0^2}{2\pi^2}
   \lim_{t \to \phi} \sum_{n=1}^\infty
   \frac{\phi^2}{\phi^2 n^2 + 4 \sin^2 \left(\frac{n \phi}{2}\right)t^2 T_0^2 r^2}
   \equiv \lim_{t \to \phi} R(t,\phi).
\end{equation}
Exchanging the limit and the series, for real values of $\phi$ and $t$, is possible because as 
a function of $t$ the \eqref{phspint3} is a uniformly convergent series of continuous functions. 
We can now apply the Theorem~\ref{th2} to the function $g(\phi) = R(t,\phi)/\phi^2$ which leads
to the following result:
\be\label{rtphi}
 R(t,\phi) \sim \frac{1}{12}\frac{T_0^2}{1+r^2 T_0^2 t^2}
    -\frac{r^2 T_0^4 t^2}{48 \pi^2\left(1+r^2 T_0^2 t^2\right)^2} \phi^2
    \pm I_f(t^2)\frac{\phi T_0^2}{2\pi^2},
\ee
where the upper sign applies to $\phi>0$ and the lower to $\phi<0$ and:
$$
 I_f(x^2)=\int_0^\infty \left[\frac{1}{y^2+4 x^2 \sin^2\left(\frac{y}{2}\right)}
    -\frac{1}{\left(1+ x^2\right) y^2}\right] {\rm d}y .
$$
Note that, again, the term $a_{-1}=0$, which excludes logarithmic terms in the asymptotic series. 
The asymptotic power series of $R(t,\phi)$ near $\phi=0$ \eqref{rtphi} has, again, a finite number of terms 
due to the fact that there are no odd powers in the expansion and $\zeta(-n)=0$ for even $n \ge 2$.
Now we can take the limit $t \to \phi$ on both sides of \eqref{rtphi} and obtain 
the asymptotic power expansion of $\langle : \wpsi^2 : \rangle_I$ about $\phi=0$, which is unique 
as it is known. This is possible because the coefficients of the right hand side of \eqref{rtphi} 
are analytic functions of $t$ and $t=\phi$ is an analytic function of $\phi$ in $\phi=0$. 
Therefore we have:
\be\label{phspint4}
  \langle : \wpsi^2 : \rangle_I = \lim_{t \to \phi} R(t,\phi) \sim 
   \frac{1}{12}\frac{T_0^2}{1+r^2 T_0^2 \phi^2} -\frac{r^2 T_0^4 \phi^4}
   {48 \pi^2\left(1+r^2 T_0^2 \phi^2\right)^2} \pm I_f(\phi^2)\frac{\phi T_0^2}{2\pi^2} .
\ee
It can be seen that the right hand side of \eqref{phspint4} can be expanded as a power series 
in $\phi$ about $\phi=0$ for $|\phi|< 1/rT_0$ by using the geometric series and thereby obtain
the full asymptotic power series of $\langle : \wpsi^2 : \rangle_I$. On the other hand, the 
\eqref{phspint4} is a resummed form of this asymptotic power series which is now suitable for 
analytic distillation. The distillation just removes the last term in \eqref{phspint4} and, going 
back to physical angular velocity with $\phi=\ii \omega/T_0$:
\be\label{phspint5}
 \langle : \wpsi^2 : \rangle = \frac{1}{12}\frac{T_0^2}{1 - r^2 \omega^2}
    -\frac{r^2 \omega^4}{48 \pi^2\left(1 - r^2 \omega^2\right)^2}.
\ee
Finally, taking advantage of relations~\eqref{eq:RotScal}, the \eqref{phspint3}
and \eqref{phspint4} yield:
\begin{equation}
\label{eq:Psi2Rot}
\langle:\wpsi^2(x):\rangle=\frac{1}{12\beta(x)^2} +\frac{\alpha(x)^2}{48\pi^2\beta(x)^2},
\end{equation}
which coincides with the expression found in the pure acceleration case \eqref{psi2-2}. 

This is indeed a remarkable and unexpected result. Starting from two considerably different 
statistical operators \eqref{accdo} and \eqref{rotdo} we have obtained, after a long mathematical 
derivation, the very same expression of the mean value of the field squared in terms of the 
Lorentz invariant thermodynamic variables. On one hand, this confirms the covariant structure
of the theory and how deep is the connection between angular velocity and acceleration in relativity.
On the other hand, it is quite surprising that in the rotating case the \eqref{eq:Psi2Rot} does 
not feature any dependence on $w^2$. This could have happened, since a scalar function like
$\langle:\wpsi^2(x):\rangle$ in principle depends on all the scalars in eq.~\eqref{lorscal} (see
discussion thereabout) and only $\alpha \cdot w$ vanishes in the pure rotating case. The 
dependence of $\langle:\wpsi^2(x):\rangle$ on the latter will be worked out in Section~\ref{newresult}.
\vspace{0.4cm}
\begin{center}
    \rule{10cm}{0.4pt}
\end{center}
\vspace{0.4cm}
Like in the pure acceleration case, we are going to calculate the thermal expectation values of the 
stress-energy tensor. To compare with expressions found in literature~\cite{AmbrusScalar,Ambrus:2014itg} 
we considered the general form of the stress-energy tensor operator of the scalar field,
which is trace-less for $m=0$ for $\xi=1/6$:
\begin{equation}\label{eq:setxi}
  \wT^{\mu\nu}_\xi= \wT_C^{\mu\nu} - \xi (\de^\mu\de^\nu-g^{\mu\nu}\square)\h\psi^\dagger\h\psi,
\end{equation}
where the canonical stress-energy tensor is given by the \eqref{canonical}. The mean 
values can be calculated with the same method of the previous section for pure acceleration: 
we study the series  associated to a complex mean value $\langle : \wT^{\mu\nu}_\xi  :\rangle_I$ 
for $\omega=-i\phi/T_0$, we take the analytic distillate and continue the result to 
physical angular velocity and lastly we take the real part. We just need to modify 
the relation~\eqref{setfinal} to account for the term proportional to $\xi$ in the 
\eqref{eq:setxi}; this is simply done by adding the following
terms to the quantities in~\eqref{deftmunueet}:
\begin{equation}\label{eq:setauxxi}
\Delta t^{\mu\nu}_\xi=-\frac{\xi}{2}\de^\mu\de^\nu \frac{1}{n^2(\tilde\beta_n \cdot \tilde \beta_n)},\quad
\Delta t_\xi=-\xi \Box \frac{1}{n^2(\tilde\beta_n \cdot \tilde \beta_n)}.
\end{equation}
From the \eqref{tildebetarot}, with $\phi = \ii\omega/T_0$, we obtain:
$$
\tilde{\beta}^\mu_n(x)=\left(
\frac{1}{T_0},\ii\frac{ x (\cos (n \phi )-1)+y \sin (n \phi )}{n},
\ii\frac{ y (\cos (n \phi )-1)-x \sin (n \phi )}{n},0 \right),
$$
\begin{equation*}
\tilde\beta_n(x) \cdot \tilde \beta_n(x) =\frac{1}{T_0^2}
    + 4 \sin^2 \left(\frac{n \phi}{2} \right) \frac{r^2}{n^2},
\end{equation*}
\begin{equation*}
\de^\mu \tilde{\beta}^\nu_n(x) = \left(
\begin{array}{cccc}
 0 & 0 & 0 & 0 \\
 0 & -\frac{\ii (\cos (n \phi )-1)}{n} & \frac{\ii \sin (n \phi )}{n} & 0 \\
 0 & -\frac{\ii \sin (n \phi )}{n} & -\frac{\ii (\cos (n \phi )-1)}{n} & 0 \\
 0 & 0 & 0 & 0 \\
\end{array}
\right),
\end{equation*}
\begin{equation*}
\begin{split}
\left(\de^\mu \de^\nu\frac{1}{\tilde\beta_n \cdot \tilde \beta_n}\right)=& \text{diag}\left( 0,
\frac{4 n^4 T_0^4 (\cos (n \phi )-1)+32 n^2 T_0^6 \left(3 x^2-y^2\right) \sin ^4\left(\frac{n \phi }{2}\right)}{\left(n^2+2 r^2 T_0^2-2 r^2 T_0^2 \cos (n \phi )\right)^3},\right.\\
&\left.-\frac{4 n^4 T_0^4 (\cos (n \phi )-1)+32 n^2 T_0^6 \left(3 y^2-x^2\right) \sin ^4\left(\frac{n \phi }{2}\right)}{\left(n^2+2 r^2 T_0^2-2 r^2 T_0^2 \cos (n \phi )\right)^3},0\right)\\
&+\frac{128 n^2 T_0^6 x y \sin ^4\left(\frac{n \phi }{2}\right)}{\left(n^2+2 r^2 T_0^2-2 r^2 T_0^2 \cos (n \phi )\right)^3}
\times\left(\begin{array}{cccc}
0 & 0 & 0 & 0 \\
0 & 0 & 1 & 0 \\
0 & 1 & 0 & 0 \\
0 & 0 & 0 & 0 \\
\end{array}\right).
\end{split}
\end{equation*}
Plugging these expressions into the~\eqref{setfinal},\eqref{deftmunueet} and \eqref{eq:setauxxi} we 
can work out every component of $\langle:\wT^{\mu\nu}_\xi:\rangle_I$ and express it as a series; the explicit forms
are given in Appendix~\ref{sec:RotSet}. Each series has the same mathematical features as the
one in~\eqref{phspint2}  and can be resummed likewise, that is by using Theorem~\ref{th2} followed
by the analytic distillation and continuation to the physical values. The details of the procedure
and the results of each components in terms of the physical rotation are given in Appendix~\ref{sec:RotSet}.
Here we report the results in terms of the thermal coefficients associated to the stress-energy tensor.
Indeed, the stress-energy tensor can be decomposed on the tetrad $\{u,\alpha,w,l \}$, which can only  depend on the Lorentz scalars \eqref{lorscal}: 
\begin{equation}
\label{Tdec}
\begin{split}
\media{:T^{\mu\nu}(x):}=&\rho\, u^\mu u^\nu -p\,\Delta^{\mu\nu} +W \, w^\mu w^\nu +{\cal A} \,\alpha^\mu \alpha^\nu
    +G^l\, l^\mu l^\nu +G\left(l^\mu u^\nu+l^\nu u^\mu\right)
    +\mathbb{A}\left(\alpha^\mu u^\nu+\alpha^\nu u^\mu\right)\\
&+G^\alpha\left(l^\mu \alpha^\nu+l^\nu \alpha^\mu\right) +\mathbb{W}\left(w^\mu u^\nu+w^\nu u^\mu\right)
    +A^w \left(\alpha^\mu w^\nu+\alpha^\nu w^\mu\right) +G^w\left(l^\mu w^\nu+l^\nu w^\mu\right),
\end{split}
\end{equation}
with $\Delta^{\mu\nu}=g^{\mu\nu}-u^\mu u^\nu$.
While a symmetric stress-energy tensor has ten degrees of freedom, the thus defined scalar 
coefficients are eleven. The redundancy of \eqref{Tdec} is owing to the fact that the tetrad
$\{u,\alpha,w,l \}$ is no longer a basis for $r=0$ because there $\alpha=l=0$. For this reason
we introduced a term $p \Delta^{\mu\nu}$ which would be redundant at finite $r$.
By projecting the stress-energy tensor onto the suitable pair of vectors (see Appendix~\ref{sec:RotSet}),
we obtain the expressions of each scalar function. The isotropic pressure is identified 
in the projection of $T^{\mu\nu}$ along the direction of $l$, which has the following continuation 
to physical values of $\omega$:
\begin{equation*}
\media{:T^{\mu\nu}:}\hat{l}_\mu \hat{l}_\nu=G^l\, l^2 -p = \frac{1-70\xi}{240\pi^2\beta^4}\left(-\alpha^2 w^2 \right)
-\left[\frac{\pi^2}{90\beta^4} -\frac{\xi}{6\beta^4} w^2 +\frac{1-6\xi}{18\beta^4} \alpha^2
    -\frac{\xi}{24\pi^2\beta^4} w^4 +\frac{19-120\xi}{1440\pi^2\beta^4} \alpha^4\right].
\end{equation*}
Since $l^2 = -\alpha^2 w^2$, the identification of $p$ and $G^l$ from the above equation 
is unambiguous. Once $p$ has been determined, all the other coefficients are obtained by 
subtraction and one finally finds:
\begin{equation}\label{setrotfinal}
\begin{split}
\rho = & \frac{\pi^2}{30 \beta^4} +\frac{4 \xi -1}{12 \beta^4}w^2 +\frac{6 \xi -1}{12 \beta^4} \alpha^2
    +\frac{4 \xi -1}{48 \pi^2 \beta^4} w^4 +\frac{60 \xi -11}{480 \pi^2 \beta^4} \alpha^4
    +\frac{270 \xi -61}{720 \pi^2 \beta^4} \alpha^2 w^2,\\
p = & \frac{\pi^2}{90\beta^4} -\frac{\xi}{6\beta^4} w^2 +\frac{1-6\xi}{18\beta^4} \alpha^2
    -\frac{\xi}{24\pi^2\beta^4} w^4 +\frac{19-120\xi}{1440\pi^2\beta^4} \alpha^4,\\
W = & \frac{2\xi-1}{12\beta^4} +\frac{2\xi-1}{48\pi^2\beta^4} w^2 +\frac{120\xi-29}{360\pi^2\beta^4} \alpha^2,\\
{\cal A}  = & \frac{1-6\xi}{12\beta^4} +\frac{1}{360\pi^2\beta^4} w^2 +\frac{1-6\xi}{48\pi^2\beta^4} \alpha^2,\\
G^l = & \frac{1-70\xi}{240\pi^2\beta^4},\\
G = & \frac{6 \xi +1}{36 \beta^4}+\frac{10 \xi -1}{240 \pi^2 \beta^4} w^2
    +\frac{30 \xi -7}{720 \pi^2 \beta^4} \alpha^2,\\
\mathbb{A} =&  0,\\
G^\alpha=& 0,\\
G^w=&A^w=\mathbb{W}=0.
\end{split}
\end{equation}
Taking into account the relations \eqref{eq:RotScal}, these results can be compared with those 
calculated in refs.~\cite{AmbrusScalar,Ambrus:2014itg} by solving the Klein-Gordon field equation 
in rotating coordinates without enforcing boundary conditions at finite radius $r$ and, in spite 
of the different methods used, precise agreement is found. 

Remarkably, by setting $w=0$ and $l=0$, as well as $\xi=0$, the resulting canonical stress-energy 
tensor is the same function of $\beta,\alpha$ found in the pure acceleration case and particularly 
the functions in \eqref{eq:ResAcc} are the same functions of $\beta^2$ and $\alpha^2$. This 
confirms the previous finding for $\langle :\wpsi^2: \rangle$ and the deep relation between 
the two examined cases. It appears, because of the covariant structure of \eqref{general}, that 
from the study of the pure rotation, one can also deduce the exact solution of thermal expectation 
values in the pure acceleration case, which is certainly a remarkable and unexpected fact. 
It should be pointed out, however, that the physical meaning of the \eqref{setrotfinal} is limited 
in the case of pure rotation; these results have been obtained by removing the inherent divergences
pertaining to the density operator~\eqref{rotdo} with the method of analytic distillation, and not 
by setting appropriate boundary conditions or other physical meaningful constraints which would make 
the operator $\widehat H - \omega \widehat J_z$ bounded from below. Indeed, the terms in 
$\alpha^2/\beta^2$, $\alpha^4/\beta^4$, $w^4/\beta^4$ and $\alpha^2 w^2/\beta^4$ are actually independent 
of $T_0$ and do not vanish in the limit $T_0 \to 0$ what would be expected in the pure rotation case for a 
well-behaved, bounded from below operator $\widehat H - \omega \widehat J_z$ in the formula \eqref{rotdo}.

\section{Thermodynamic equilibrium with rotation and acceleration: a new result}
\label{newresult}

So far we have shown that this method reproduces the exact results known or virtually known in literature,
which were obtained by solving field equations in curvilinear coordinates. However, the main advantage
of the proposed method is that it allows to obtain the exact expressions for thermal expectation
values even when the solutions in curvilinear coordinates are not known or are difficult to handle,
for instance when both acceleration and rotation in the equations~\eqref{thvort} and \eqref{alphaw}
are non-vanishing. To prove it, as a case study, we compute the thermal expectation value 
$\langle : \wpsi^2 : \rangle$ in a thermodynamic equilibrium state with both rotation and
acceleration along the $z$-axis, that is with thermal vorticity:
\begin{equation*}
    \varpi^{\mu\nu}=\frac{\omega}{T_0}(g^{\mu 1} g^{\nu 2}-g^{\mu 2}g^{\nu 1})
    - \frac{a}{T_0}(g^{\mu 0}g^{\nu 3}-g^{\mu 3} g^{\nu 0}) \qquad \implies \qquad
    \beta^\mu=\frac{1}{T_0}(1+a z,-\omega y,\omega x, t a).
\end{equation*}
Contrary to the previous cases, $\langle : \wpsi^2 : \rangle$ may now depend
on the Lorentz scalar $(\alpha \cdot w)$, which is non-vanishing:
$$
  \alpha \cdot w = - \frac{ a \omega}{T^2_0},
$$  
whereas the other Lorentz scalars it may depend on (see equation \eqref{lorscal} and foregoing
discussion) turn out to be:
\begin{align}\label{lorscal2}
    &\beta^2=\frac{1}{T_0^2}\left((1+az)^2-a^2t^2-r^2\omega^2\right), &&\alpha^2= 
    -\frac{a^2 \left((1+az)^2-a^2 t^2\right)+r^2 \omega ^4}{\beta^2 T_0^4}, \\ \nonumber
    &w^2=-\frac{\omega ^2 \left((1+ a z)^2-a^2 t^2+a^2 r^2(a z+1)^2\right)}{T_0^4\beta^2} ,
\end{align}
where $r^2 = x^2 + y^2$.

We start by computing the $\tilde\beta$ vector according to the equation \eqref{tildebetaf}. For this 
purpose, one can take advantage of the following relation: 
$$
\varpi^\mu_\alpha \varpi^\alpha_\nu = \frac{1}{T_0^2}\left(\begin{matrix}
a^2 &0 &0 &0 \\ 0 &-\omega^2 &0 &0 \\ 0 &0 &-\omega^2 &0\\ 0 &0 &0 &a^2
\end{matrix}\right)
$$
which returns a diagonal matrix. The sum \eqref{tildebetaf} can then be readily split into odd and 
even powers, leading to an analytic expression involving both hyperbolic and trigonometric functions:
\begin{equation*}
\begin{split}
    \widetilde{\beta}(\varpi)=&\left(\frac{(a z+1) \sin \left(\frac{a}{T_0}\right) 
   - \ii a t \left(\cos \left(\frac{a }{T_0}\right)-1\right)}{a },-y \sinh
   \left(\frac{\omega }{T_0}\right)-\ii x \left(\cosh \left(\frac{\omega }{T_0}\right)-1\right),\right. \\
   &\left.x \sinh \left(\frac{\omega }{T_0}\right)-\ii y \left(\cosh \left(\frac{\omega }{T_0}\right)-1\right),
   \frac{a t \sin \left(\frac{a }{T_0}\right) - \ii (a z+1) \left(\cos \left(\frac{a }{T_0}\right)-1\right)}{a }\right).
\end{split}
\end{equation*}
After some lengthy but simple calculations, the expectation value can be obtained (see eq.~\eqref{psi2}
and following) for imaginary acceleration and vorticity $a/T_0\rightarrow -\ii \Phi$ and 
$\omega/T_0\rightarrow-\ii \phi$:
\begin{equation*}
\begin{split}
   \langle:\wpsi^2(x):\rangle_I =\frac{1}{2\pi^2}\sum_{n=1}^\infty\frac{1}{n^2 \widetilde{\beta}(-n\varpi)
   \cdot\widetilde{\beta}(-n\varpi)}=\frac{1}{8 \pi ^2}\sum_{n=1}^\infty 
   \frac{\Phi^2}{ r^2 \Phi ^2 \sin ^2\left(\frac{n \phi }{2}\right)+\sinh ^2\left(\frac{n \Phi }{2}\right) 
   \left(t^2 \Phi^2+\left(\frac{1}{T_0}-i \Phi  z\right)^2\right)}.
\end{split}
\end{equation*}

The series as it stands, with $\phi$ and $\Phi$ real, is uniformly convergent. On the other hand, for 
general complex values of $\phi$ and/or $\Phi$ the series diverges and an analytic distillation procedure 
can be carried out by limiting $\phi$ and $\Phi$ to be real, much the same way as for equilibrium 
with rotation. It is thus convenient to write $\Phi$ and $\phi$ in polar coordinates $\Phi=\xi \cos\theta$ 
and $\phi=\xi\sin\theta$ and focus on the analytic distillation of the variable $\xi$ whose limit $\xi \to 0$
corresponds to both $\phi$ and $\Phi$ vanishing. We first write the series in a form which is suitable 
for the application of the Zagier's Theorem \ref{th2}:
\begin{equation}\label{psi2-3}
    \langle:\wpsi^2(x):\rangle_I = \sum_{n=1}^{\infty} \lim_{ B \rightarrow B(\xi,\theta)} \lim_{C \rightarrow C(\xi,\theta)}
    \frac{1}{8\pi^2} \frac{\xi^2 \cos^2\theta}{r^2 \, C^2\sin^2\left(\frac{n\xi\sin\theta}{2}\right)+ 
    B\sinh^2\left(\frac{n\xi\cos\theta}{2} \right)}  \; \; ,
\end{equation}
where we introduced the auxiliary parameters $B$ and $C$:
\be\label{bfunct}
B(\xi,\theta) = t^2 \xi^2\cos^2\theta^2+\left(\frac{1}{T_0}-i \xi\cos\theta  z\right)^2,
\qquad
C(\xi,\theta) = \xi \cos\theta.
\ee
Except for $B=0$\footnote{Since $T_0>0$, it is guaranteed that $B\neq 0$.}, the exchange of series with 
limits is legitimate because the \eqref{psi2-3} is a uniformly convergent series of continuous functions. 
Now we can apply the Theorem \ref{th2} to the function
$$ 
g(\xi,\theta) = \frac{1}{8\pi^2}\sum_{n=1}^{\infty} \frac{1}{r^2 \, C^2\sin^2\left(\frac{n\xi\sin\theta}
{2}\right)+ B \sinh^2\left(\frac{n\xi\cos\theta}{2} \right)}
$$
and obtain the asymptotic power series of $g$ for $\xi\rightarrow 0$:
\be\label{gasym}
g(\xi,\theta) \sim \frac{1}{12 \left(B\cos^2\theta  +C^2 r^2\sin^2\theta\right)}\frac{1}{\xi ^2}-
\frac{C^2 r^2 \sin^4\theta -B \cos^4\theta}{48 \pi ^2 \left(B \cos^2 \theta +C^2 r^2 \sin^2\theta\right)^2}
\pm\frac{I(\theta,B,C)}{\xi }
\ee
being:
$$
I(\theta,B,C)=\int_0^\infty\di y\left[\frac{1}{8 \pi ^2 \left(B \sinh ^2\left(\frac{1}{2} y \cos\theta \right)
+C^2 r^2 \sin ^2\left(\frac{1}{2} y \sin\theta\right)\right)}-\frac{1}{2\pi^2 y^2 \left(B\cos^2\theta+C^2 
r^2\sin^2\theta\right)}\right].
$$
Since $B(\xi,\theta)$ and $C(\xi,\theta) = \xi \cos \theta$ are analytic functions of $\xi$, and 
the coefficients of the powers of $\xi$ in the \eqref{gasym} are analytic functions 
of $\xi$ in $\xi=0$, we can obtain the full resummed asymptotic expression of $g(\xi,\theta)$ 
about $\xi=0$ by taking the limits of $B$ and $C$ to $B(\xi,\theta)$ and $C(\xi,\theta)$ 
respectively:
\begin{equation*}
\begin{split}
g(\xi,\theta) &\sim \frac{1}{12 \left(\left(t^2 \xi^2\cos^2\theta
   +\left(\frac{1}{T_0}-i \xi\cos\theta  z\right)^2\right)\cos^2\theta  
   +\xi^2 r^2\cos^2\theta \sin^2\theta\right)}\frac{1}{\xi ^2}+\\
   &-\frac{\xi^2r^2\cos^2\theta  \sin
   ^4\theta -\left(t^2 \xi^2\cos^2\theta
   +\left(\frac{1}{T_0}-i \xi\cos\theta  z\right)^2\right) \cos^4\theta}
   {48 \pi ^2 \left(\left(t^2 \xi^2\cos^2\theta +\left(\frac{1}{T_0}-i \xi\cos\theta  
   z\right)^2\right) \cos^2 \theta +\xi^2r^2\cos^2\theta  \sin^2\theta\right)^2}\pm\frac{I(\xi,\theta)}{\xi }
   \end{split}
\end{equation*}
much like in the rotation case in Section \ref{rotation}. We are now in a position to apply the 
distillation process to the above function; we obtain:
\begin{equation*}
\begin{split}
\dist_{0}\xi^2\cos^2\theta g(\xi,\theta)&=\frac{\cos^2\theta}{12 \left(\left(t^2 \xi^2\cos^2\theta
   +\left(\frac{1}{T_0}-i \xi\cos\theta  z\right)^2\right)\cos^2\theta  +\xi^2r^2\cos^2\theta \sin^2\theta\right)}+\\
   &-\frac{\xi^2r^2\cos^2\theta  \sin
   ^4\theta -\left(t^2 \xi^2\cos^2\theta
   +\left(\frac{1}{T_0}-i \xi\cos\theta  z\right)^2\right) \cos^4\theta}{48 \pi ^2 \left(\left(t^2 \xi^2\cos^2\theta
   +\left(\frac{1}{T_0}-i \xi\cos\theta  z\right)^2\right) \cos^2 \theta +\xi^2r^2\cos^2\theta  \sin^2\theta\right)^2}\xi^2\cos^2\theta .
   \end{split}
\end{equation*}
Going back to the variables $\Phi$ and $\phi$ and analitically continuing to real acceleration and 
angular velocity we get:
\begin{equation}
\langle:\wpsi^2(x):\rangle = \frac{T_0^2}{12 \left((1+az)^2-a^2t^2-r^2 \omega^2\right)}-
 \frac{a^2 \left((1+az)^2-a^2t^2 \right) +r^2 \omega ^4}
 {48 \pi ^2 \left((1+az)^2-a^2t^2-r^2 \omega ^2\right)^2},
\end{equation}
which, by using the equations \eqref{lorscal2}, can be rewritten as:
\begin{equation}
    \langle:\wpsi^2(x):\rangle=\frac{1}{12\beta(x)^2}+\frac{\alpha(x)^2}{48\pi^2}.
\end{equation}
The result is the same as in the case of pure acceleration \eqref{psi2-2} and pure rotation
\eqref{eq:Psi2Rot}; there is no dependence on the possible argument $\alpha\cdot w$. Therefore, 
the equation \eqref{psi2-2} applies to any kind of global equilibrium for the massless scalar 
field, that is for any tensor $\varpi$. Achieving this conclusion by solving the field equations 
would have been extremely hard.

\section{Summary} 

In summary, we have derived a general exact form of the phase space distribution function and 
the thermal expectation values of local operators of the free quantum scalar field in the most
general case of thermodynamic equilibrium in Minkowski space-time, with the four-temperature being
a Killing field, that is including rotation and acceleration. 
The presented derivation does not make use of the solutions of the Klein-Gordon equations in 
curvilinear coordinates but it is just based on the plane wave expansion of the field in Minkowski 
space-time and it is therefore suitable for the general case including both rotation and 
linear acceleration. The crucial steps of the derivation are a factorization of the general density 
operator \eqref{general} using Poincar\'e group algebra and an iterative method to obtain the 
thermal expectation values of quadratic combinations of creation and annihilation operators. The 
general form of the phase space distribution function has been written as a formal series including 
all quantum corrections to the classical term. We have studied the series in two major cases of 
non-trivial equilibrium, the pure acceleration and the pure rotation and compared with exact
known results obtained solving Klein-Gordon equation in Rindler and rotating coordinates respectively. 
Furthermore, we have obtained a completely new result, that is the thermal expectation value 
of the massless scalar field squared in the case of global equilibrium with both acceleration 
and rotation. The result shows that this quantity does not depended on the scalar product 
of acceleration and rotation vectors $A \cdot \omega$. It should be emphasized that
it would have been extremely hard to obtain the same result by means of the solutions of
Klein-Gordon equation in curvilinear coordinates.
In the case of pure acceleration, the iterative method introduces undesired non-analytic terms 
to be subtracted, while in the rotation case the absence of boundary conditions at finite radius
$ < 1/\omega$ implies unavoidable physical divergences.
The resummation of the series in these two cases required the introduction of a new operation on 
complex functions, defined as analytic distillation, in order to extract the analytic part in
the pure acceleration case and analytically continue the finite solution for imaginary angular
velocity in the pure rotation case. In the former case, the method of analytic distillation leads 
to expressions in the massless case which are the same as obtained in Rindler coordinates and, 
remarkably, automatically vanish at the Unruh temperature; mathematically, a new class of complex 
polynomials is generated by this physics problem, which all vanish for $z=2\pi \ii$. In the case 
of pure rotation, on the other hand, it should be pointed out that the analytic distillation and 
continuation provides a finite result which is in agreement with analytic calculations in 
literature, but with limited physical meaning.

\section*{Acknowledgments}

We are greatly indebted to D. Dorigoni for letting us know about Zagier's theorem and his
work on Lambert series. We warmly thank D. Basile and D. Dorigoni for illuminating discussions
about asymptotic series and the extraction of an analytic part. We thank F. Colomo and 
C. Dappiaggi for equally useful discussions.

\appendix

\section{Recurrence formula}\label{recurrence}

We want to work out ${\sf \Lambda}^{-s}\tilde b (\varpi)$
for a non-negative integer $s$ and ${\sf \Lambda} = \exp [\varpi: {\sf J}/2]$ being
$\varpi$ complex in general. By using the eqs.~\eqref{btilde} and the expansion 
of ${\sf \Lambda}$, we have:
$$
{\sf \Lambda}^{-s} \tilde b(\varpi) = \sum_{h=0}^\infty \left( \frac{-s \varpi: {\sf J}}{2}
\right)^h \frac{1}{h!} \sum_{k=0}^\infty \frac{1}{(k+1)!}\left( \frac{\varpi: {\sf J}}{2} 
\right)^k b .
$$
This expression can be worked out as follows:
\begin{align*}
 & \sum_{h=0}^\infty \left( \frac{-s \varpi: {\sf J}}{2}
\right)^h \frac{1}{h!} \sum_{k=0}^\infty \frac{1}{(k+1)!}\left( \frac{\varpi: {\sf J}}{2} 
\right)^k b = \sum_{h,k=0}^\infty (-s)^h \frac{1}{h! (k+1)!} \left( \frac{\varpi: {\sf J}}{2}
\right)^{h+k} b \\
 & = \sum_{n=0}^\infty \left( \frac{\varpi: {\sf J}}{2} \right)^{n} b \sum_{h=0}^n (-s)^h 
  \frac{1}{h!(n+1-h)!} = \sum_{n=0}^\infty \frac{1}{(n+1)!}
   \left( \frac{\varpi: {\sf J}}{2} \right)^{n} b \sum_{h=0}^n (-s)^h \frac{(n+1)!}{h!(n+1-h)!} ,
\end{align*}
where we set $n=h+k$. We can write the last obtained expression with binomial coefficients:
\be\label{recurr-a}
 {\sf \Lambda}^{-s}\tilde b (\varpi) = \sum_{n=0}^\infty \frac{1}{(n+1)!}
   \left( \frac{\varpi: {\sf J}}{2} \right)^{n} b \sum_{h=0}^n (-s)^h \binom{n+1}{h}.
\ee
Now, since:
$$
 \sum_{h=0}^n (-s)^h \binom{n+1}{h} = \sum_{h=0}^{n+1} (-s)^h \binom{n+1}{h} - (-s)^{n+1}
 = (1-s)^{n+1} + s (-s)^n
$$
the eq.~\eqref{recurr-a} can be written as:
\begin{align*}
 {\sf \Lambda}^{-s}\tilde b (\varpi) &= \sum_{n=0}^\infty \frac{1}{(n+1)!}
  (1-s) \left( \frac{(1-s) \varpi: {\sf J}}{2} \right)^{n} b + 
  \sum_{n=0}^\infty \frac{1}{(n+1)!} s \left( \frac{(-s) \varpi: {\sf J}}{2} \right)^{n} b
  \\
  &= (1-s) \tilde b((1-s)\varpi) + s \tilde b(-s\varpi) .
\end{align*}
%

\section{Solutions of the homogeneous equation}\label{solhom}

We provide an instance of a non-analytic solution of equation \eqref{homog} for $\phi=0$. 
In general, if ${\sf \Lambda}= \exp[- \ii \phi : {\sf J}]$ and there is a non-vanishing vector 
$v(\phi)$ such that:
\be\label{bv}
  \tilde b = ({\sf I}-{\sf \Lambda})v(\phi)
\ee
where $\tilde b$ is given by the \eqref{btilde} with $\phi = \ii \varpi$, then:
$$
  H(p,p^\prime) = G(p^\prime,\phi) \exp[v(\phi) \cdot p]
$$
solves the equation \eqref{homog} for any function $G(p^\prime,\phi)$. Note that, since $\tilde b$ 
is real for real $\sf \Lambda$, $v(\phi)$ must be real as well. Since:
$$
 \media{\wad{p} \wa{p^\prime}}_\phi =  
  \frac{1}{Z} \Tr \left( \exp [ - b \cdot \wP - \ii \phi : \wJ] \wad{p} \wa{p^\prime} \right)
  = \frac{1}{Z} \Tr \left( \wad{p^\prime} \wa{p} \exp [ - b \cdot \wP + \ii \phi : \wJ] \right)^*
  = \media{\wad{p^\prime} \wa{p}}_{-\phi}^*
$$
taking into account that $v$ is real, we have:
$$
  G(p^\prime,\phi) \exp[ v(\phi) \cdot p] = G(p,-\phi)^* \exp[ v(-\phi) \cdot p^\prime].
$$
This functional equation, applying to any $p,p^\prime$ is solved by the combination:
\be\label{solution}
  H(p,p^\prime) = F(\phi) \exp[v(\phi)\cdot p + v(-\phi) \cdot p^\prime]
\ee
with $F(\phi)^* = F(-\phi)$.

If ${\sf \Lambda}$ is a rotation, then $\tilde b = b$ (see Section~\ref{rotation}) and since
$b$ is proportional to the time unit vector $\hat t$, there is no real vector $v(\phi)$ 
solving the \eqref{bv}. If, on the other hand, ${\sf \Lambda}$ is a pure boost, say along
the $z$ axis, then we have that \eqref{bv} has non-trivial solutions. By using the equation
\eqref{tildebacc} and the definition of $x_0= (0,0,0,-1/a)$ (see Section~\ref{sec:accel})
it can be checked that: 
$$
v(\phi)=\frac{1}{T_0}(0,0,0,-\frac{1}{\phi})
$$ 
is a solution of the \eqref{bv} with $\phi = \ii a/T_0$. In this case, it turns
out that, according to the above general proof, $H(p,p^\prime)$ is non-analytic in $\phi=0$.

\section{Analytic distillation for pure acceleration and the Unruh effect}\label{distillac}

Just like the thermal expectation value of the canonical stress-energy tensor~\eqref{canonical},
any local operator which is quadratic in the field and its derivatives has a thermal
expectation value given by a four-momenta integral of the Wigner function~\eqref{cantens} or
equivalently of the phase space distribution function and its derivatives~\eqref{cantens2}. 
Note that the appearance of the real part in eq.~\eqref{cantens2} is a consequence of 
the hermiticity of the stress-energy tensor and, more in general, of any physical 
observable. As already pointed out in the main text, our knowledge of the distribution 
function is limited to~\eqref{phspacec4} which holds for imaginary vorticity.
From~\eqref{phspacec4}, the thermal expectation value of local operators is obtained by 
considering integrals of this sort:
\begin{equation}\label{eq:momint}
\int\frac{\di^3 \p}{\varepsilon}\,p^{\mu_1}\cdots p^{\mu_N}\, \de^{\nu_1}\cdots \de^{\nu_M}\,
    \e^{- n \tilde \beta_n \cdot p} = \de^{\nu_1}\cdots \de^{\nu_M}\,
    \frac{\partial}{\partial \tilde\beta_n^{\mu_1}}
    \ldots \frac{\partial}{\partial \tilde\beta_n^{\mu_N}}
    \frac{(-1)^N}{n^N} \int\frac{\di^3 \p}{\varepsilon} \; \e^{- n \tilde \beta_n \cdot p}.
\end{equation}
These integrals must be computed for imaginary vorticity, and the real part, implied 
by hermiticity, must be taken only after the analytic distillation and the continuation 
back to real vorticity.

We want to study such integrals for the pure acceleration case, with imaginary acceleration 
$a/T_0 = -\ii \phi$ and demonstrate explicitly that - in the massless neutral case - after analytic 
distillation and continuation they give rise to expressions vanishing at the Unruh temperature 
$T_0 = a/2\pi$. This might be a consequence of the distribution function~\eqref{phspacec4} being
independent of $p$ and $x$ when evaluated at the Unruh temperature, as shown in Section~\ref{sec:accel}.
Since that is a general feature of the distribution function, we can expect that the expressions for the
massive and charged case will also be vanishing, as dictated by the Unruh effect.
Moreover, it is sufficient to calculate the integrals~\eqref{eq:momint} in $x=0$ because the
whole dependence on $x$ of any tensor field is fully constrained (see discussion after eq.~\eqref{xindep}). 

By using eq.~\eqref{eq:MasslesMomInt} and eq.~\eqref{tildebetaaccn}, it can be realized that 
the equation \eqref{eq:momint} in $x=0$ gives rise to linear combinations of the following series:
\begin{equation}
\label{eq:sinhseries}
S_{2m+2}(\phi)=\sum_{n=1}^\infty  \frac{\phi^{2m+2}}{\sinh^{2m+2}(n\phi/2)},\quad
S^{(1)}_{2m+2}(\phi)=\sum_{n=1}^\infty  \frac{\phi^{2m+2}\sinh(n \phi)}{\sinh^{2m+2}(n\phi/2)}.
\end{equation}
As shown in the text specifically for $S_2(\phi)$, these are series of analytic functions 
if ${\rm Re} \, \phi \ne 0$ and they are uniformly convergent in the same domain, so they 
define analytic functions when ${\rm Re}\, \phi \neq 0$. Conversely, these series are divergent 
whenever $\phi$ is purely imaginary. Nevertheless, we can obtain asymptotic power series about 
$\phi=0$. 

As far as the series $S_{2m+2}^{(1)}(\phi)$ in~\eqref{eq:sinhseries} is concerned, it can be shown
that its asymptotic expansion about $\phi=0$ is given by a series of odd powers of $\phi$, hence its
contribution to the physical expectation values, after distillation, continuation, and extraction of the real
part, vanishes. 

We are then left with $S_{2m+2}(\phi)$ only. With $\phi > 0$ real and positive, we can use the 
Theorem~\ref{th2} to obtain an asymptotic power series of the function $G_{2m+2}(\phi)$:
\begin{equation*}
G_{2m+2}(\phi)\equiv\phi^{-2m-2} S_{2m+2}(\phi)=\sum_{n=1}^\infty  \frac{1}{\sinh^{2m+2}(n\phi/2)}
    \equiv\sum_{n=1}^\infty f(n \phi).
\end{equation*}
By using the generalized version of the Bernoulli polynomials $B^{(m)}_n(t)$ defined by~\cite{Norlund1924}
\begin{equation*}
    \left(\frac{x}{\e^x-1}\right)^m \e^{tx} =\sum_{n=0}^\infty\frac{B^{(m)}_n(t)}{n!}x^n,
\end{equation*}
the function $f$ can be written as a power series about $\phi=0$:
\begin{equation}
\label{espansione 1/sinh}
\begin{split}
f(\phi)=&\frac{1}{\sinh^{2m+2}\left(\frac{\phi}{2}\right)}
    = 2^{2m+2} \phi^{-2m-2}\left(\frac{\phi}{\e^\phi-1}\right)^{2m+2} \e^{(m+1)\phi} \\
    =& 2^{2m+2}\sum_{n=0}^\infty\frac{B^{(2m+2)}_n(m+1)}{n!}\phi^{n-2m-2}
    = 2^{2m+2}\sum_{n=-2m-2}^\infty\frac{B^{(2m+2)}_{2m+2+n}(m+1)}{(2m+2+n)!}\phi^n,
\end{split}
\end{equation} 
which converges for any complex $\phi$ with $|\phi|<2\pi$~\cite{Norlund1924}. Owing to the parity 
of the function $f$, the coefficient $B^{(2m+2)}_n(m+1)$ appearing in~\eqref{espansione 1/sinh} must be 
vanishing for odd $n$. The conditions of the Theorem~\ref{th2} for its application to the function
$G_{2m+2}(\phi)$ are fulfilled and we have, in its notation:
\begin{equation*}
a_{-1}=2^{2m+2}\frac{B^{(2m+2)}_{2m+2-1}(m+1)}{(2m+2-1)!}=0
\end{equation*}
as well as
\begin{equation*}
\begin{split}
I_f=&\int_{0}^{\infty} \left[\frac{1}{\sinh^{2m+2}\left(\frac{\phi}{2}\right)}
    -2^{2m+2}\sum_{n=-2m-2}^{-2}\frac{B^{(2m+2)}_{2m+2+n}(m+1)}{(2m+2+n)!}\phi^n\right]{\rm d} \phi\\
=&\lim_{\epsilon\to 0^+}\int_{\epsilon}^{\infty} \left[\frac{1}{\sinh^{2m+2}\left(\frac{\phi}{2}\right)}
    -2^{2m+2}\sum_{n=-2m-2}^{-2}\frac{B^{(2m+2)}_{2m+2+n}(m+1)}{(2m+2+n)!}\phi^n\right]{\rm d} \phi = (-1)^{m+1}\frac{4}{m+2},
\end{split}
\end{equation*}
where in the last step  we iteratively used~\cite[Formula 1 \S 1.4.5, p. 146]{book:PrudVol1}.
Therefore, the asymptotic expansion of $G_{2m+2}(\phi)$ reads:
$$
   G_{2m+2}(\phi) \sim  (-1)^{m+1}\frac{4}{m+2}\frac{1}{\phi} 
   + 2^{2m+2}\sum_{n=-2m-2}^\infty \frac{B^{(2m+2)}_{2m+2+n}(m+1)}{(2m+2+n)!} \zeta(-n) \phi^n.
$$
Since $a_{-1}=0$ and the function $f$ in eq.~\eqref{espansione 1/sinh} is analytic in the region
${\rm Re} \, \phi > 0$, we can extend the above expansion to complex $\phi$ in the same region
according to the discussion following Theorem \ref{th2} (see also figure~\ref{fig:paths} in Section~\ref{math}). Notice that the above sum effectively stops at $n=0$ because, for positive $n$, 
the Zeta function is non-vanishing only when $n$ is odd, but in this case the generalized Bernoulli polynomial coefficients vanish. Then, by taking advantage of the relation between the Riemann Zeta 
function and the Bernoulli numbers
$$
\zeta(2n)=\frac {(-1)^{n+1}B_{2n}(2\pi )^{2n}}{2(2n)!},
$$
the asymptotic expansion of the hyperbolic series can be written as:
\begin{equation}
\label{eq:SAsym}
\begin{split}
S_{2m+2}(\phi)\sim &-2^{2m+1}\sum_{k=0}^{m+1}(2\pi\ii)^{2k}\frac{B_{2k}}{(2k)!}
	\frac{B_{2m+2-2k}^{(2m+2)}(m+1)}{(2m+2-2k)!} \, \phi^{2m+2-2k}+(-1)^{m+1}\frac{4}{m+2}\phi^{2m+1}.
\end{split}
\end{equation}
For instance, in the cases $m=0,1$ the asymptotic expansions are:
\begin{equation*}
\begin{split}
S_2(\phi)\sim &\frac{2\pi^2}{3} + \frac{\phi^2}{6}-2\phi,\\
S_4(\phi)\sim &\frac{8\pi^4}{45}-\frac{4\pi^2}{9}\phi^2-\frac{11}{90}\phi^4
    +\frac{4}{3}\phi^3.
\end{split}
\end{equation*}
Repeating this procedure in the region where Re$\phi<0$, we find that the asymptotic expansion of
$S_{2m+2}(\phi)$ is the same as in eq.~\eqref{eq:SAsym} with the opposite sign of the term proportional
to $\phi^{2m+1}$, which is hence removed by the distillate.

Now we can extract the analytic distillate of 
the function $S_{2m+2}(\phi)$ and continue 
it to imaginary values of $\phi$, that is to a real acceleration. Like in Section~\ref{sec:accel}
the result of distillation are the polynomials:
\be\label{bpoly}
\dist_{0} S_{2m+2}(\phi) =
\mathcal{B}_{2m+2}(\phi) \equiv -2^{2m+1}\sum_{k=0}^{m+1}(2\pi\ii)^{2k}\frac{B_{2k}}{(2k)!}
	\frac{B_{2m+2-2k}^{(2m+2)}(m+1)}{(2m+2-2k)!} \, \phi^{2m+2-2k} ,
\ee
which are to be continued to imaginary $\phi$.

We now prove that the polynomials \eqref{bpoly} have two zeroes in $\phi=\pm 2\pi\ii$ for any 
$m\geq 0$, hence their analytic continuation for real accelerations $a/T_0 = -\ii \phi$ vanish
precisely at the Unruh temperature. If $\phi=2\pi\ii$ is a zero of $\mathcal{B}_{2m+2}$ then the following identity must be true:
\begin{equation}
\label{eq:IdentZeros}
\mathcal{B}_{2m+2}(2\pi\ii)
=\sum_{k=0}^{m+1}\frac{B_{2k}\,B_{2m+2-2k}^{(2m+2)}(m+1)}{(2k)!(2m+2-2k)!}=0.
\end{equation}
To prove the identity~\eqref{eq:IdentZeros}, we start showing that for every integer $m> 0$
the generalized Bernoulli polynomials have a zero in $m$:
\begin{equation}
\label{eq:ZerosBern}
B^{(2m+1)}_{2m}\left(m\right)=0.
\end{equation}
With $z\in\mathbb{C}$ and $\alpha,\beta\geq 0$ integers, the following ratio of Euler Gamma
functions has an exact power series representation~\cite{LukeVol1}:
\begin{equation}\label{gamser}
\frac{\Gamma(z+\alpha)}{\Gamma(z-\beta)}=\sum_{k=0}^{\alpha+\beta} \frac{(\alpha+\beta)!}{k!}
    \frac{B_{\alpha+\beta-k}^{(1+\alpha+\beta)}(\alpha)}{(\alpha+\beta-k)!} z^{k}.
\end{equation}
Then, for $m> 0$ and setting $\alpha=\beta=m$, the above equation becomes:
\begin{equation*}
\frac{\Gamma(z+m)}{\Gamma(z-m)}=\sum_{k=0}^{m} \frac{(2m)!}{k!(2m-2k)!} B_{2m-k}^{(2m+1)}(m)
\, z^{k}.
\end{equation*}
If we evaluate the previous expression in $z=0$ and we use the reflection formula for the Gamma 
function, we have:
\begin{equation*}
\frac{\Gamma(m)}{\Gamma(-m)}=\Gamma(m)\Gamma(m+1)\frac{\sin(m\pi)}{\pi}=0.
\end{equation*}
For $z=0$ the sum \eqref{gamser} contains just one term, hence we have:
\begin{equation*}
\frac{\Gamma(m)}{\Gamma(-m)}=B^{(2m+1)}_{2m}\left(m\right)=0,
\end{equation*}
which proves the equation \eqref{eq:ZerosBern}.

To proceed towards the proof of \eqref{eq:IdentZeros}, let us now consider the function
\begin{equation}
\label{auxf}
f(x)=\left(\frac{x}{\e^x-1}+\frac{x}{2}\right)\left(\frac{x\,\e^{x/2}}{\e^x-1}\right)^{2m+2},
\end{equation}
which involves the generators of the Bernoulli polynomials and has
a power series expansion:
\begin{equation}\label{svilf}
f(x)=\sum_{k=0}^\infty d_k\, x^k\, ,
\end{equation}
which is convergent for $|x|\leq 2\pi$. We are going to show that, for any integer $m \ge 0$:
\begin{equation}
\label{eq:devenId}
d_{2m+2}=\sum_{l=0}^{m+1} \frac{B_{2l}}{(2l)!} \frac{B^{(2m+2)}_{2m+2-2l}(m+1)}{(2m+2-2l)!}=0
\end{equation}
which proves the identity~\eqref{eq:IdentZeros}. Consider the even terms of \eqref{svilf},
i.e. $d_{2k}$. Splitting the product in \eqref{auxf} as
\begin{equation*}
f(x)=\frac{x}{\e^x-1}\left(\frac{x\,\e^{x/2}}{\e^x-1}\right)^{2m+2}
	+\frac{x}{2}\left(\frac{x\,\e^{x/2}}{\e^x-1}\right)^{2m+2},
\end{equation*}
it can be realized that the second term is odd and does not contribute to the $d_{2k}$
while the first term gives rise to the generalized Bernoulli polynomials:
\begin{equation*}
\frac{x}{\e^x-1}\left(\frac{x\,\e^{x/2}}{\e^x-1}\right)^{2m+2}=
\left(\frac{x}{\e^x-1}\right)^{2m+3} \e^{(m+1)x}=
\sum_{k=0}^\infty \frac{1}{k!}B^{(2m+3)}_{k}(m+1) x^k.
\end{equation*}
From this expansion and eq.~\eqref{eq:ZerosBern} it follows that for every integer $m\geq 0$ 
the coefficient $d_{2m+2}$ vanishes:
\begin{equation*}
 d_{2m+2}=\frac{1}{(2m+2)!}B^{(2m+3)}_{2m+2}(m+1)=0.
\end{equation*}
To prove that $d_{2m+2}$ equates the right hand side of \eqref{eq:devenId}, we need to write the 
full power series expansion of $f$ in~\eqref{svilf}, which can be obtained with the product of
the series of the two functions enclosed in the brackets of \eqref{auxf}. The first factor in 
the right hand side of \eqref{auxf} can be expanded as:
\begin{equation}
\label{eq:Cauchy1}
\frac{x}{\e^x-1}+\frac{x}{2}=\sum_{i=0}^\infty \frac{B_{2i}}{(2i)!} x^{2i};
\end{equation}
while the second factor is the generating function of the generalized Bernoulli polynomials of 
order $2m+2$:
\begin{equation}
\label{eq:Cauchy2}
\left(\frac{x\,\e^{x/2}}{\e^x-1}\right)^{2m+2}=\sum_{j=0}^\infty \frac{B^{(2m+2)}_{2j}(m+1)}{(2j)!} x^{2j}.
\end{equation}
Hence, by making the Cauchy product of~\eqref{eq:Cauchy1} and~\eqref{eq:Cauchy2}
in the \eqref{auxf} we obtain:
\begin{equation*}
f(x)=\sum_{k=0}^\infty \sum_{l=0}^k \frac{B_{2l}}{(2l)!} \frac{B^{(2m+2)}_{2k-2l}(m+1)}{(2k-2l)!} x^{2k}.
\end{equation*}
Equating this expression with the \eqref{svilf}, we obtain the expression~\eqref{eq:devenId}
for the vanishing coefficient $d_{2m+2}$, which finally proves the identity~\eqref{eq:IdentZeros}.

Taking advantage of the~\eqref{eq:IdentZeros}, the polynomial~\eqref{bpoly} can also be 
written as:
\begin{equation*}
\begin{split}
 \mathcal{B}_{2m+2}(\phi) = &\sum_{i=0}^m \frac{2^{2m+1}B_{2i+2}}{(2i+2)!}
    \frac{B_{2m-2i}^{(2m+2)}(m+1)}{(2m-2i)!}\phi^{2(m-i)}\left(\phi^{2i+2}-(2\pi\ii)^{2i+2}\right),
\end{split}
\end{equation*}
which manifestly shows that $\mathcal B_{2m+2}$ vanishes at $\phi=\pm2\pi\ii$.

\section{Stress-energy tensor for a massless scalar field in Rindler coordinates}
\label{exactrind}

We calculate the thermal expectation value of the stress-energy tensor of the massless scalar 
field with the density operator \eqref{accdo} by using the solutions of the Klein-Gordon equation in 
Rindler coordinates, in order to compare with the results \eqref{eq:ResAcc}. The task will 
be accomplished by taking advantage of calculations presented in refs.~\cite{,Becattini:2017ljh,Becattini:2019poj}.

The symmetries of \eqref{accdo} constrain the mean value to be of the form \eqref{eq:setAcc}. 
On the other hand, from quantum field theory and using the equation of motion for a neutral 
massless field, the canonical stress-energy tensor may be written as:
\begin{equation*}
    \wT^{\mu\nu}=\nabla^\mu\wpsi\nabla^\nu\wpsi-\frac{1}{2}g^{\mu\nu}
    \nabla_\rho\wpsi\nabla^\rho\wpsi=\nabla^\mu\wpsi\nabla^\nu\wpsi-\frac{1}{4}g^{\mu\nu}\square\wpsi^2.
\end{equation*} 
We kept the covariant derivatives for future convenience. Combining this expression with 
\eqref{eq:setAcc} the following expression are obtained:
\begin{subequations}
\label{setcontract}
\begin{align}
\rho_R=&\langle:u_\mu u_\nu \wT^{\mu\nu}:\rangle_R
    =\langle:\left(u\cdot\nabla \wpsi\right)^2:\rangle_R-\frac{1}{4}\square\langle:\wpsi^2:\rangle_R,\\
p_R-\frac{{\cal A}_R}{3}\alpha^2=&-\frac{1}{3}\Delta_{\mu\nu}\langle:\wT^{\mu\nu}:\rangle_R
    =\frac{1}{3}\left(\rho_R+\frac{1}{2}\square\langle:\wpsi^2:\rangle_R\right),\\
-p_R\alpha^2+{\cal A}_R(\alpha^2)^2=&\alpha_\mu\alpha_\nu\langle:\nabla^\mu\wpsi\nabla^\nu\wpsi:\rangle_R
    -\frac{\alpha^2}{4}\square\langle:\wpsi^2:\rangle_R,
\end{align}
\end{subequations}
where $u^\mu$ is the four-velocity $\beta^\mu/\sqrt{\beta^2}$ and $\alpha^\mu=\sqrt{\beta^2} A^\mu$
where $A^\mu$ is the acceleration field, see Section~\ref{intro}. Note that in this section the subscript 
$R$ specifies that normal ordering applies to Rindler's creation and annihilation operators and not 
to Minkowski's, like it is understood in the rest of the paper. Since they have non-trivial Bogoliubov 
transformations, the two normal-ordering are not equivalent.

From the four-temperature in \eqref{accfields} it ensues (Cartesian components):
\begin{align*}
    u^\mu=\frac{1}{k}(z',0,0,t), && A^\mu=\frac{1}{k^2}(t,0,0,z'),
\end{align*}
where $z'=1+az$ and $k=\sqrt{(z'^2-t^2)}$. 
The thermal expectation values of the fields can be calculated from the solution of the Klein-Gordon
equation in Rindler coordinates. In fact, this calculation has already been partially carried out 
in~\cite{Becattini:2019poj}, where it was found that:
\begin{align*}
  \langle:\left(u\cdot\nabla \wpsi\right)^2:\rangle_R=\frac{\pi^2}{30\beta^4},   
  &&\square\langle:\wpsi^2:\rangle_R=\frac{\alpha^2}{3\beta^4},
\end{align*}
with $\alpha^2=\alpha_\mu \alpha^\mu$. With these results, the \eqref{setcontract} becomes:
\begin{subequations}
\label{setcontract2}
\begin{align}
    \rho_R=&\frac{\pi^2}{30\beta^4}-\frac{\alpha^2}{12\beta^4},\\
    p_R-\frac{{\cal A}_R}{3}\alpha^2=&\frac{\pi^4}{90\beta^4}+\frac{\alpha^2}{36\beta^4},\\
    -p_R\alpha^2+{\cal A}_R(\alpha^2)^2=
    &\alpha_\mu\alpha_\nu\langle:\nabla^\mu\wpsi\nabla^\nu\wpsi:\rangle-\frac{(\alpha^2)^2}{12\beta^4},\label{setcontract2-last}
\end{align}
\end{subequations}
and we just need to calculate one more thermal expectation value. Before doing that, it is useful 
to remind some useful relations in Rindler coordinates. 
The relation between Cartesian and Rindler coordinates (called here $\tau$ and $\xi$) is given by:
\begin{align*}
    t=\frac{\e^{a\xi}}{a}\sinh(a\tau), && z'=\frac{\e^{a\xi}}{a}\cosh(a\tau).
\end{align*}
Using these expressions, we can show that:
\begin{equation*}
    \frac{\mathrm{d}x^\mu}{\mathrm{d}\xi}=\frac{\e^{2a\xi}}{a} A^\mu.
\end{equation*}
Also, in Rindler coordinates, $k^2=a^{-2}\e^{2a\xi}$ and $\beta^2=\e^{-2a\xi} T_0^{-2}$.
Now we can work out the thermal expectation value in eq.~\eqref{setcontract2-last}:
\begin{equation}
\label{eq:lastmean}
\langle:\left(\alpha_\mu\nabla^\mu\wpsi\right)^2:\rangle_R
    =\beta^2\langle:\left(A_\mu\nabla^\mu\wpsi\right)^2:\rangle_R
    =\beta^2 a^2\e^{-4\xi a}\left\langle:\left(\frac{\mathrm{d}\wpsi}{\mathrm{d}\xi}\right)^2:\right\rangle_R
    =-\alpha^2 \e^{-2\xi a}\left\langle:\left(\frac{\mathrm{d}\wpsi}{\mathrm{d}\xi}\right)^2:\right\rangle_R.
\end{equation}
In the right Rindler wedge, the scalar massless field is given by~\cite{Crispino:2007eb}:
\begin{equation*}
\wpsi(\tau,\xi,\mathbf{x}_T)^{(R)}=\int_0^\infty\mathrm{d}\omega\int\mathrm{d}^2\mathbf{k}_T
    u_{\omega,\mathbf{k}_T} a^{(R)}_{\omega,\mathbf{k}_T}
        +u^*_{\omega,\mathbf{k}_T}{a^\dagger}^{(R)}_{\omega,\mathbf{k}_T}
\end{equation*}
where the eigenfunctions $u_{\omega,\mathbf{k}_T}$ read:
\begin{equation*}
u_{\omega,\mathbf{k}_T}^{(R)}=\sqrt{\frac{\sinh\left(\frac{\pi\omega}{a}\right)}{4\pi^4 a}}
    K_{\frac{i\omega}{a}}\left(\frac{k_T\e^{a\xi}}{a}\right)
    \e^{i\mathbf{k}_T\cdot\mathbf{x}_T} \e^{-i\omega\tau},
\end{equation*}
$K_n(z)$ being the modified Bessel function of the second kind and $\mathbf{k}_T$
and $\mathbf{x}_T$ are the transverse momenta and coordinates. Using the mean value
of the number operator in the right Rindler wedge \cite{Becattini:2017ljh}:
\begin{equation*}
\langle {a^\dagger}^{(R)}_{\omega,\mathbf{k}_T}a^{(R)}_{\omega',\mathbf{k}'_T}\rangle
    =\frac{\delta(\omega-\omega')\delta^2(\mathbf{k}_T-\mathbf{k}'_T)}{\e^{\frac{\omega}{T_0}}-1}
\end{equation*}
and the formula:
\begin{equation*}
    \frac{\partial K_\nu(x)}{\partial x}=-\frac{1}{2}\left(K_{\nu+1}(x)+K_{\nu-1}(x)\right),
\end{equation*}
we obtain the integrals:
\begin{equation*}
    \langle:\left(\frac{d\wpsi}{d\xi}\right)^2:\rangle_R=\frac{\e^{2a\xi}}{4\pi^3a}\int _0^\infty\di\omega\frac{\sinh\left(\frac{\pi\omega}{a}\right)}{\e^{\frac{\omega}{T_0}}-1}\sum_{\pm}
    \int_0^\infty\di k_T \, k_T^3 K_{\pm1+i\frac{\omega}{a}}\left(\frac{k_T\e^{\xi a}}{a}\right)K_{\pm1-i\frac{\omega}{a}}\left(\frac{k_T\e^{\xi a}}{a}\right),
\end{equation*}
where the sum runs over all possible combinations of signs $\pm$ in the order of the Bessel
functions. The integration over $\mathbf{k}_T$ can be done by means of the known 
integral~\cite{gradshteyn2007}:
\begin{equation*}
\begin{split}
\int_0^\infty\mathrm{d}x \; x^{-\lambda}K_\mu(ax)K_\nu(ax)=&
    \frac{2^{-2-\lambda}a^{\lambda-1}}{\Gamma(1-\lambda)}
    \Gamma\left( \frac{1-\lambda+\mu+\nu}{2}\right)
    \Gamma\left(\frac{1-\lambda-\mu+\nu}{2}\right)\\
    &\times\Gamma\left( \frac{1-\lambda+\mu-\nu}{2}\right)\Gamma\left(\frac{1-\lambda-\mu-\nu}{2}\right),
\end{split}
\end{equation*}
which applies to ${\rm Re}\, a>0$ and ${\rm Re}\, \lambda<1-|{\rm Re}\,\mu|-|{\rm Re}\,\nu|$. After 
integrating over $\omega$ we get:
\begin{equation*}
    \langle:\left(\alpha_\mu\nabla^\mu\wpsi\right)^2:\rangle_R=\frac{\pi^2\alpha^2}{90\beta^4}-
    \frac{\left(\alpha^2\right)^2}{9\beta^4},
\end{equation*}
and, putting it in the \eqref{setcontract2}, we have:
\begin{align*}
\rho_R=&\frac{\pi^2}{30\beta^4}-\frac{\alpha^2}{12\beta^4},\\
p_R-\frac{{\cal A}_R}{3}\alpha^2=&\frac{\pi^4}{90\beta^4}+\frac{\alpha^2}{36\beta^4},\\
-p_R\alpha^2+{\cal A}_R(\alpha^2)^2=&-\alpha^2\left(\frac{\pi^2}{90\beta^4}-\frac{\alpha^2}{9\beta^4}
        -\frac{\alpha^2}{12\beta^4}\right);
\end{align*}
so that, finally:
\begin{equation*}
     \rho_R=\frac{\pi^2}{30\beta^4}-\frac{\alpha^2}{12\beta^4},\quad
     p_R=\frac{\pi^2}{90\beta^4}+\frac{\alpha^2}{18\beta^4},\quad
     {\cal A}_R=\frac{1}{12\beta^4}.
\end{equation*}

The last step is to calculate the stress-energy tensor with subtraction of the Minkowski 
vacuum, which was implied in the normal-ordering used throughout the paper.
As has been discussed in detail in ref.~\cite{Becattini:2017ljh}, the subtraction of 
Minkowski vacuum corresponds to the subtraction of the normally-ordered Rindler expression
at the Unruh temperature which is such that $T_U^2=-A_\mu A^\mu/4\pi^2$.
So, for the Minkowski normally-ordered quantities (the subscript is dropped) we have:
\begin{align*}
     \rho=&\frac{\pi^2}{30\beta^4}-\frac{\alpha^2}{12\beta^4}-\frac{11\alpha^4}{480\pi^2\beta^4},\\
     p=&\frac{\pi^2}{90\beta^4}+\frac{\alpha^2}{18\beta^4}+\frac{19\alpha^4}{1440\pi^2\beta^4},\\
     {\cal A}=&\frac{1}{12\beta^4}+\frac{\alpha^2}{48\pi^2\beta^4} .
\end{align*}
which agree with the equations \eqref{eq:ResAcc}.

\section{Analytic distillation of the stress-energy tensor for pure rotation}\label{sec:RotSet}

Here we provide some details about how to derive the stress-energy tensor of the neutral scalar 
field in the pure rotation case, as it appears in~\eqref{setrotfinal}. For an imaginary
angular velocity to temperature ratio $\phi=\ii\omega/T_0$, by using the equations~\eqref{setfinal},
\eqref{deftmunueet} and \eqref{eq:setauxxi} one can write each component of the stress-energy tensor 
as a series. Defining the components $\Theta$ with:
\begin{equation*}
\media{:\h{T}_\xi^{\mu\nu}(x):}_I=\sum_{n=1}^\infty
    \frac{T_0^4\phi^4}{\pi^2 \left(n^2\phi^2 + 4 \sin^2 \left(\frac{n \phi}{2}\right) r^2 T_0^2\phi^2\right)^3}
 \Theta^{\mu\nu}_n(x),
\end{equation*}
one obtains, for the non-vanishing ones:
\begin{equation*}
\begin{split}
\Theta^{00}_n(x)=& (n\phi)^2(4\xi +2)+\cos(n\phi)\left[(n\phi)^2(1-4\xi)+2 (8\xi -1) r^2 T_0^2\phi^2\right]
    -\cos(2n\phi)(4\xi -1) r^2 T_0^2\phi^2-(12\xi -1) r^2 T_0^2\phi^2,\\
\Theta^{xx}_n(x)=& (n\phi)^2(1-2\xi)+ 2\cos(n\phi)\left[(n\phi)^2\xi
    +(4\xi-1)T_0^2\phi^2\left(x^2-3 y^2\right)\right] -2 (3\xi -1) T_0^2\phi^2\left(x^2-3 y^2\right)+\\
    &+2 T_0^2\phi^2\left[x y (\sin(2n\phi)-2 \sin(n\phi))-\xi \left(x^2-3 y^2\right) \cos(2n\phi)\right], \\
\Theta^{yy}_n(x)=& (n\phi)^2(1-2\xi)+ \cos(n\phi)\left[2(n\phi)^2\xi
    -2(4\xi-1)T_0^2\phi^2\left(3 x^2-y^2\right)\right] +2 (3\xi -1)T_0^2\phi^2\left(3 x^2-y^2\right)+\\
    &+2 T_0^2\phi^2\left[x y (2 \sin(n\phi)-\sin(2n\phi))+\xi\left(3 x^2-y^2\right) \cos(2n\phi)\right], \\
\Theta^{zz}_n(x)=& (n\phi)^2(2-4\xi)-\cos(n\phi)\left[(n\phi)^2(1-4\xi)-2(4\xi-1)r^2T_0^2\phi^2\cos(n\phi)
    +2 (8\xi-1) r^2 T_0^2\phi^2\right]+8\xi r^2 T_0^2\phi^2, \\
\Theta^{0x}_n(x)=&  \Theta^{x0}_n(x)=-2\ii  (n\phi) (T_0\phi)\left[x (\cos(n\phi)-1)-2 y \sin(n\phi)\right], \\
\Theta^{0y}_n(x)=&  \Theta^{y0}_n(x)= 2 \ii (n\phi)(T_0\phi)\left[-2 x \sin(n\phi)-y (\cos(n\phi)-1)\right], \\
\Theta^{xy}_n(x)=&  \Theta^{yx}_n(x)=  4 T_0^2\phi^2 \sin^2\left(\frac{n \phi }{2}\right)
    \left[4 x y (1-2\xi+2\xi  \cos(n\phi))+(x-y) (x+y) \sin(n\phi)\right].
\end{split}
\end{equation*}
As in the case of the series~\eqref{phspint2}, the poles of these series cover the whole
complex plane. However, we can still obtain a proper asymptotic power series if we restrict
$\phi$ to take on only real values. Notice that the variable $\phi$ never appears alone
but always multiplied by either $n$ or $T_0$. So,
we can replace $\phi$ with an auxiliary variable $t$ every time it appears with a coefficient 
$T_0$. By means of this replacement, new functions $R^{\mu\nu}(n\phi,t,x)$ are defined and 
the thermal expectation values of the stress-energy tensor can be obtained by taking the limit:
\begin{equation*}
\media{:\h{T}_\xi^{\mu\nu}(x):}_I=\sum_{n=1}^\infty \lim_{t\to\phi} R^{\mu\nu}(n\phi,t,x),
\end{equation*}
as it was done in the main text for $\media{:\wpsi^2:}_I$. Indeed, these series are of the same kind
of that in eq.~\eqref{phspint3}. Also, it can be realized that for real values of $\phi,\,t$ and $x$ 
they are all uniformly convergent series of continuous functions and, consequently, we can exchange 
the limit and the sum:
\begin{equation}
\label{eq:ZagierR}
\media{:\h{T}_\xi^{\mu\nu}(x):}_I=\lim_{t\to\phi}\sum_{n=1}^\infty R^{\mu\nu}(n\phi,t,x).
\end{equation}
The form~\eqref{eq:ZagierR} fulfills the requirements of the Theorem~\ref{th2}, which can then
be used to obtain the asymptotic expansion of the thermal expectation values around $\phi=0$.
Thereafter, like for the eq.~\eqref{phspint3}, the asymptotic expansion is used to obtain the 
analytic distillate of~\eqref{eq:ZagierR} in $\phi=0$. Finally, the stress-energy tensor
components are obtained restoring the physical angular velocity and taking the real part:
\begin{equation*}
\begin{split}
\media{:T^{00}(x):}&= \frac{\pi^2}{30} \frac{4 \gamma^2-1}{3}\gamma^4 T_0^4
     +\frac{1-4\xi}{12} \frac{6\gamma^4-\gamma^2(24\xi 
+1)+12\xi-2}{3-12\xi}\gamma^4 T_0^2\omega^2+\\
     &+\frac{4\xi-1}{48\pi^2} \frac{120 \gamma^6-3 \gamma^4 (240\xi 
+17)+\gamma^2 (720\xi -58)
         -120\xi +19}{30(1-4\xi)}\gamma^4 \omega^4 , \\
\media{:T^{xx}(x):}=& \frac{\pi^2}{90} \gamma^6 T_0^4 
+\frac{\xi}{6}\frac{15\left(\gamma^2+4\xi-1\right)
         -4\pi^2T_0^2\left(x^2-3 y^2\right)}{60\xi}\gamma^6T_0^2\omega^2+\\
     &-\frac{\xi}{24\pi^2}\frac{3 \left(5 \gamma^4+\gamma^2 (30\xi 
-9)-20\xi +4\right)
         +10\pi^2 T_0^2\left(3\gamma^2+12\xi -2\right) \left(x^2-3 
y^2\right)}{30 \xi}\gamma^6\omega^4+\\
     &+\frac{30 \gamma^4+3 \gamma^2 (60\xi -13)-60\xi +11}{1440 \pi^2} 
\left(x^2-3 y^2\right)\gamma^6 \omega^6, \\
\media{:T^{yy}(x):}=& \frac{\pi^2}{90} \gamma^6 T_0^4 
+\frac{\xi}{6}\frac{15\left(\gamma^2+4\xi-1\right)
         -4\pi^2T_0^2\left(y^2-3 x^2\right)}{60\xi}\gamma^6T_0^2\omega^2+\\
     &-\frac{\xi}{24\pi^2}\frac{3 \left(5 \gamma^4+\gamma^2 (30\xi 
-9)-20\xi +4\right)
         +10\pi^2T_0^2 \left(3\gamma^2+12\xi -2\right) \left(y^2-3 
x^2\right)}{30 \xi}\gamma^6\omega^4+\\
  &-\frac{30 \gamma^4+3 \gamma^2 (60\xi -13)-60\xi 
+11}{1440\pi^2}\left(3 x^2-y^2\right)\gamma^6\omega^6, \\
\media{:T^{zz}(x):}=& \frac{\pi^2}{90}\gamma^4 T_0^4
     +\frac{4\xi-1}{12}  \frac{\gamma^2 (24\xi -5)-12\xi +2}{3(4\xi-1)} 
\gamma^4 T_0^2 \omega^2+\\
&+\frac{1-4\xi}{48\pi^2}\frac{15\gamma^4(48\xi-11)+\gamma^2(154-720\xi)+120\xi-19}{30(4\xi-1)}\gamma^4\omega^4,\\
\media{:T^{0x}(x):}=&  \media{:T^{x0}(x):}
     = -\frac{2\pi^2 \gamma^6}{45} y T_0^4 \omega
     -\frac{\left(3 \gamma^2-1\right)\gamma^6}{18} y T_0^2 \omega^3
     +\frac{\left(5 \gamma^2-4\right) \gamma^8}{60 \pi^2} y \omega^5, \\
\media{:T^{0y}(x):}=&  \media{:T^{y0}(x):}
     = \frac{2\pi^2 \gamma^6}{45} x T_0^4 \omega
     +\frac{\left(3 \gamma^2-1\right)\gamma^6}{18} x T_0^2 \omega^3
     -\frac{\left(5 \gamma^2-4\right) \gamma^8}{60 \pi^2} x \omega^5, \\
\media{:T^{xy}(x):}=& \media{:T^{yx}(x):}
     = -\frac{2\pi^2 \gamma^6}{45} T_0^4 x y \omega^2
     -\frac{\left(3 \gamma^2+12\xi -2\right)\gamma^6}{18}  x y T_0^2 
\omega^4+\\
     &+\frac{\left(30 \gamma^4+3 \gamma^2 (60\xi -13) -60\xi 
+11\right)\gamma^6}{360 \pi^2}x y \omega^6,
\end{split}
\end{equation*}
where $\gamma = u^0 = 1/\sqrt{1-r^2\omega^2}$.
The thermal coefficients defined by the decomposition~\eqref{Tdec}, i.e.
\begin{equation*}
\begin{split}
\media{:T^{\mu\nu}(x):}=&\rho\, u^\mu u^\nu -p\,\Delta^{\mu\nu} +W \, w^\mu w^\nu
    +{\cal A} \,\alpha^\mu \alpha^\nu  +G^l\, l^\mu l^\nu +G\left(l^\mu u^\nu+l^\nu u^\mu\right)
    +\mathbb{A}\left(\alpha^\mu u^\nu+\alpha^\nu u^\mu\right)\\
&+G^\alpha\left(l^\mu \alpha^\nu+l^\nu \alpha^\mu\right) +\mathbb{W}\left(w^\mu u^\nu+w^\nu u^\mu\right)
    +A^w \left(\alpha^\mu w^\nu+\alpha^\nu w^\mu\right) +G^w\left(l^\mu w^\nu+l^\nu w^\mu\right),
\end{split}
\end{equation*}
can be obtained by projecting the stress-energy tensor along the vectors of the tetrad:
\begin{equation*}
\begin{split}
u^\mu=&\gamma  \left( 1,-y \omega ,x \omega ,0 \right),\quad
\alpha^\mu= \left( 0,-\frac{\gamma  x \omega ^2}{T_0},-\frac{\gamma  y \omega^2}{T_0},0\right),\\
w^\mu=& \left( 0,0,0,\frac{\gamma  \omega }{T_0} \right),\quad
l^\mu= \left( \frac{\gamma  \left(\gamma^2-1\right) \omega^2}{T_0^2},-\frac{\gamma^3 y \omega ^3}{T_0^2}
    ,\frac{\gamma^3 x \omega^3}{T_0^2},0\right).
\end{split}
\end{equation*}
Finally, the scalar functions in the projections are written in terms of the scalars of~\eqref{eq:RotScal}, 
namely:
\begin{equation*}
\beta^2(x)=\frac{1}{T_0^2\gamma^2},\quad \alpha^2(x)=-(\gamma^2-1)\frac{\omega^2}{T_0^2},
\quad w^2(x)=-\gamma^2\frac{\omega^2}{T_0^2}.
\end{equation*}
The final expressions are reported in the equations~\eqref{setrotfinal}.
\clearpage

\bibliographystyle{apsrev4-1}
\bibliography{Biblio}

\end{document}